\documentclass[a4paper,11pt]{article}
\pdfoutput=1 

\usepackage{jheppub} 

\usepackage[T1]{fontenc} 

\usepackage[dvipsnames]{xcolor}
\usepackage{verbatim}
\usepackage{array}
\usepackage{multirow}
\usepackage{booktabs}
\usepackage{subfig}
\usepackage{tikz}
\usetikzlibrary{calc,through,backgrounds,patterns,trees,positioning,arrows,shapes,chains,%
	decorations.pathmorphing,decorations.markings,decorations.pathreplacing,fadings}

\usepackage{pgfplots}

\usepgfplotslibrary{fillbetween}    
\pgfplotsset{compat=1.10}

\newcommand{\CL}{\mathcal{L}}
\newcommand{\CP}{\mathcal{P}}

\newcommand{\CN}{\mathcal{N}}

\newcommand{\CR}{\mathcal{R}}

\newcommand{\ads}{\mbox{AdS}}

\def\Tr{\mathrm{Tr}~}

\def\imm{\mathrm{Im}}
\def\real{\mathrm{Re}}

\def\ie{\textit{i.e.} }
\def\eg{\textit{e.g.} }

\title{\boldmath AC conductivities of a holographic Dirac semimetal}

\author[a]{Gianluca Grignani,}
\author[a]{Andrea Marini,}
\author[b]{Lorenzo Papini}
\author[a]{and Adriano-Costantino Pigna}

\affiliation[a]{Dipartimento di Fisica e Geologia, Universit\`a di Perugia,\\
	I.N.F.N. Sezione di Perugia,\\
	Via Pascoli, I-06123 Perugia, Italy}
\affiliation[b]{Dipartimento di Fisica e Astronomia ``G. Galilei'', Universit\`a di Padova  \\
	I.N.F.N. Sezione di Padova, \\
	Via Marzolo 8, I-35131 Padova, Italy}

\emailAdd{grignani@pg.infn.it}
\emailAdd{andrea.marini@pg.infn.it}
\emailAdd{lorenzo.papini@pd.infn.it}
\emailAdd{pigna@pg.infn.it}

\abstract{We use the AdS/CFT correspondence to compute the AC conductivities for a (2+1)-dimensional system 
	of massless fundamental fermions coupled to (3+1)-dimensional Super Yang-Mills theory at strong coupling.
	We consider the system at finite charge density, with a constant electric field along the defect
	and an orthogonal magnetic field.
	The holographic model we employ is the well studied D3/probe-D5-brane system. 
	There are two competing phases in this model: 
	a phase with broken chiral symmetry favored when the magnetic field dominates over 
	the charge density and the electric field and a chirally symmetric phase in the opposite regime.
	The presence of the electric field induces Ohm and Hall currents, which can be straightforwardly computed by means of the 
	Karch-O'Bannon technique.
	Studying the fluctuations around the stable configurations in linear response theory, we are able to derive the full frequency dependence of
	longitudinal and Hall conductivities in all the regions of the phase space.}

\begin{document} 
	\maketitle
	\flushbottom


\section{Introduction}

The AdS/CFT correspondence is a duality between low-energy effective theories of string theory and supersymmetric gauge theories. It was originally conjectured as an equality between Type IIB Supergravity on $\ads_5 \times S^5$ background and the supersymmetric $\CN = 4$ $SU(N_c)$ Super Yang-Mills quantum field theory in the limit $N_c \to \infty$ and $\lambda \to \infty$ with $\lambda = g_{YM} \, N_c^2$ \cite{Maldacena:1997re, Gubser:1998bc, Witten:1998qj}. Since the quantum field theory is in a strong coupling regime when the gravity one is a low-energy effective theory, the correspondence is also a strong/weak coupling duality. For this reason it has been proven to be a formidable tool to evaluate relevant physical quantities for strongly coupled field theories by means of the gravity dual ones. 

A (2+1)-dimensional semimetal is an example of a  physical system for which the AdS/CFT correspondence seems to be particularly well suited. 
This can be motivated using graphene as a representative of semimetals. Although some of its properties can be studied through perturbative approaches, there are some theoretical evidences, like the small Fermi velocity ($v_F\sim c/300$), and some experimental ones \cite{Crossno:2016,b92bfba555b644c5a82fba0fc93cd45b} which suggest that interactions in graphene may be strong. 
If this would be the case an accurate description
of the physics in graphene would require a non perturbative approach and in this scenario the AdS/CFT correspondence represents the best analytical tool at our disposal.

\begin{figure}[!htb]
	\centering
	\begin{tikzpicture}
	\def\lungDt{9};
	\def\largDt{2};
	\def\altDcu{1.5};
	\def\altDcd{2.5};
	\coordinate (D3a) at (-.5*\lungDt,0);
	\coordinate (D5a) at (-1,0);
	\fill[Cerulean!70,opacity=.8,path fading=south] (D5a) --  ++(2,2)  -- ++(0,-\altDcd) -- ++(-2,-2) -- cycle; 
	\fill[Green!50,opacity=.8] (D3a) --  ++(\lungDt,0)  -- ++(2,\largDt) -- ++(-\lungDt,0) -- cycle; 
	\fill[Cerulean!70,opacity=.8] (D5a) --  ++(0,\altDcu)  -- ++(2,2) -- ++(0,-\altDcu) -- cycle; 
	\draw[Green!60!black,line width=3](D3a) --  ++(\lungDt,0) node[above left=0pt and 2pt] {\footnotesize D3} -- ++(2,\largDt);
	\draw[Cerulean!50!black,line width=1.5] ($(D5a)-(0,\altDcd)$) --  ++(0,\altDcu+\altDcd) node[anchor=north west]{\footnotesize D5} -- ++(2,2);
	\draw[Blue,line width=2,line cap=round] (D5a) -- +(2,2) node[near end] (A) {};
	\fill[white,path fading=north] (-3,-3) rectangle +(8,2.5);
	\node (T) at (3,3) {\footnotesize (2+1)-dim defect};
	\draw [blue!70!black,line width=1,->] (A) to[in=-120,out=0] (T); 
	\end{tikzpicture}	
	\caption{Illustration of the D3/probe-D5-brane setup.}
	\label{fig:d3d5}	
\end{figure}
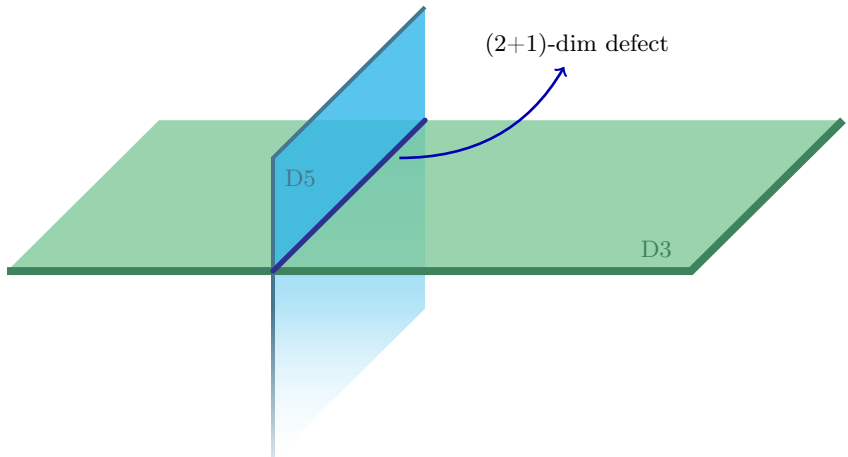

The study of Dirac semimetals with holographic techniques can be approached using either bottom-up  
(see for instance \cite{Seo:2016vks,Rogatko:2017tae,Rogatko:2017svr} for recent applications) or, as we do in this paper, top-down models, based on 
D-branes constructions. In particular we consider the well studied D3/probe-D5-brane system,
where the D5-probes intersect the D3-branes on a (2+1)-dimensional defect, as depicted in Figure~\ref{fig:d3d5}.
This turns out to be a good holographic model to describe the physics governing charge carriers in graphene, as 
can be seen by considering the field theory dual of the system, which
consists in fundamental matter particles living on the (2+1)-dimensional defect and
interacting through $\CN=4$ Super Yang-Mills degrees of freedom in $3+1$ dimensions \cite{DeWolfe:2001pq, Karch:2000gx, Erdmenger:2002ex}.
Taking zero asymptotic separation between the D3 and D5-branes corresponds to having 
massless fundamental particles on the defect. This is exactly what we want for graphene, 
where charge carriers are known to be massless  at the kinetic level.
Thus in the dual string theory picture we can interpret the (2+1)-dimensional brane intersection 
as the holographic realization of the graphene layer.

The geometry of the D$5$-brane probes at the boundary is fixed to be
$\ads_4 \times S^2$.
If no external scale is introduced, it turns out that the whole 
geometry of the D$5$-brane worldvolume is actually given by $\ads_4 \times S^2$ and this 
gives a global $SO(3)\times SO(3)$ symmetry to the theory. 
When an external magnetic field $B$ is turned on, the D$5$-brane geometry changes:
the probe brane pinches off before reaching the Poincar\'e horizon (Minkowski embedding) and the $SO(3)\times SO(3)$ 
symmetry is broken to a $SO(3)\times U(1)$.
In the dual field theory this can be viewed as a chiral symmetry breaking  
due to the formation of a fermion-antifermion condensate \cite{Filev:2007gb,Filev:2009xp}. 
The introduction of either finite charge density $\rho$ or finite temperature $T$ opposes this condensation, giving rise to a more interesting 
phase diagram, with a transition from the phase with broken symmetry to the symmetric one as the ratios $\rho/B$ or $T/B$ increases \cite{Evans:2010hi}. 
At zero temperature the chiral symmetry breaking transition happens at $\rho/B=\sqrt{7}$ and it turns out to be a BKT phase transition \cite{Kaplan:2009kr}. For small $T$ it is of second order nature \cite{Jensen:2010ga} with changing $\rho/B$ and
for small $\rho$ it is of first order with changing $T/B$.
When the charge density is small but finite the D5-brane geometry still breaks the chiral symmetry but in a different fashion compared to
the zero charge case, since this time the D$5$-brane worldvolume reaches the horizon (Black Hole embedding). 
This can be simply understood in the holographic picture where charge carriers are represented by F$1$-strings, which, having higher tension with 
respect to D5-branes, pull the latter down to the horizon.

The D3/probe-D5-brane setup was also used to model double monolayer semimetal systems 
formed by two parallel sheets of a semimetal separated by an insulator \cite{Evans:2013jma,Grignani:2014vaa,Grignani:2016npu}. 
In this case one has to consider two stacks of probe branes (a stack of D5 and one of anti-D5) to represent holographically the two semimetal layers.
The presence of the two layers introduces another parameter in the model, namely the separation between them, and a new channel for
the chiral symmetry breaking, driven by the condensation between a fermion on one layer and an antifermion on the other one.

The aim of this paper is to derive the AC conductivity matrix for a (single layer) (2+1)-dimensional semimetal, such as graphene, using 
the holographic D3/probe-D5-brane model. 
In particular we will consider the D3/probe-D5 system with mutually perpendicular electric and magnetic field at finite charge 
density. The presence of the electric field $E$ is necessary in order to have non trivial Ohm and Hall currents.
When $E$ is different from zero, the on-shell action for the probe branes becomes
generally complex at a critical locus, usually called singular shell, on the brane worldvolume 
and in order to avoid this one has to turn on the Ohm and Hall currents and suitably fix their values in terms of the 
parameters of the system (\eg $E$, $B$, $\rho$, $\dots$) \cite{Karch:2007pd}.
The same system we consider, also with finite temperature, was studied before in \cite{Evans:2011tk} and the values of the 
DC currents were derived imposing the reality condition on the on-shell action.

The holographic derivation of the AC conductivity matrix for systems involving probe D$p$-branes, similar to the one we are considering,
was addressed by several papers in literature. 
For example, in \cite{Hartnoll:2009ns} probe flavour D$p$-branes in the context of a neutral Lifshitz-invariant quantum critical theory were considered and the AC conductivity with non trivial charge density, temperature and electric field and vanishing magnetic field was obtained. 
The authors of Ref.~\cite{Das:2010yw} studied probe D$p$-branes rotating in an internal sphere direction and derived the AC conductivity of the system considering nonzero electric field and charge density and vanishing temperature.
The results of \cite{Hartnoll:2009ns, Das:2010yw} are both compatible with a finite temperature regime, as suggested
for instance by the presence of a finite peak at low frequency. In \cite{Das:2010yw} this is a consequence of the fact that there is an induced horizon by the rotation of the D$p$-branes and therefore there is an effective nonzero induced Hawking temperature proportional to the frequency of rotation. 
In the system we consider we expect to find, at least in some regimes, a similar physics 
since when the singular shell is outside the Poincar\'e horizon it plays the role of an induced horizon, 
resulting in a finite effective temperature.

The strategy we use to evaluate the AC conductivity matrix is the following. We focus on the linear response regime and we 
fluctuate gauge and scalar fields upon a fixed background. 
Then we solve the equations of motion of the action which rules the dynamic of the system \ie the DBI action. 
We obtain the equations of motion for the gauge field fluctuations $A^{(1)}_a (t,r) = e^{-i \omega t} a_a (r)$ 
and we solve them numerically. The AC conductivities in the linear response regime can be evaluated using the Kubo formula
\begin{equation}
\sigma_{ij}(\omega)= - i \frac{G_{j_i j_j}^R(\omega)}{\omega}\, ,
\end{equation}
where $G_{j_i j_j}^R$ is the retarded current-current Green's function.
Using the holographic dictionary this can be computed as
\begin{equation}
\sigma_{ij}=\lim\limits_{r\rightarrow 0} -\frac{i}{\omega}\, \frac{a'_{i}(r)}{a_j(r)}\, ,
\end{equation}
in terms of the $r$-dependent part of the gauge field fluctuations, $a_i(r)$.

\medskip

The paper is structured as follows. In Section~\ref{sec:model} we will describe in detail the holographic model we consider. We will show its action, discuss how the currents are naturally fixed by reality conditions which must be imposed on the on-shell Routhian and we show the phase diagram of the system. 
In Section~\ref{sec:fluctuations} we derive the effective action for the fluctuations for the D$3$/D$5$ system in a very general framework, considering both scalar fields and gauge field fluctuations.
Section~\ref{sec:conductivities} is devoted to the computation of AC conductivity matrices for all the relevant phases of the system. 
For each of these phases we show some plots of Ohm and Hall conductivities. 
We conclude with Section~\ref{sec:discussion} where we discuss the obtained results.

\section{The holographic model}
\label{sec:model}

The holographic model we consider is the D3/probe-D5-brane system. 
In this section we briefly summarize the setup and the allowed 
configurations for this system. These will constitute the 
background configurations around which we fluctuate in order to study the 
conductivities.

\subsection{D-brane setup}

We start by considering a stack of $N$ D3-branes, which as usual we replace with the $\ads_5\times S^5$
geometry that they generate in the near horizon limit. In the coordinate system we use, the $\ads_5\times S^5$ metric reads
\begin{equation}
\label{eq:ads5}
ds^2=G_{\mu\nu}dx^\mu dx^\nu
=\frac{1}{r^2}\left(-dt^2+dx^2+dy^2+dz^2+dr^2\right)+d\psi^2+\sin^2\psi \,d^2\Omega_2+\cos^2\psi \,d^2\tilde{\Omega}_2\, ,
\end{equation}
where $d^2\Omega_2=d\theta^2+\sin^2\theta d\varphi^2$ and $d^2\tilde \Omega_2=d\tilde\theta^2+\sin^2\tilde \theta d\tilde \varphi^2$ 
are the metrics of two 2-spheres, $S^2$ and $\tilde S^2$.  
The AdS boundary is located at $r=0$ and the Poincar\'e horizon at $r=\infty$.

We now embed $N_5$ D5-branes as probes in this background. We choose 
$\sigma^a =(t,x,y,r,\theta_1 , \varphi_1)$ as D5 worldvolume coordinates and we also
allow the D5-branes to have a non-trivial profile along $\psi$. The choice of the
embedding is summarized in Table~\ref{tab:D5coords}.
\begin{table}[!ht]
	\centering
	\begin{tabular}{ccccccccccc}
		\toprule
		& $t$ & $x$ & $y$ & $z$ & $r$ & $\psi$ & $\theta_1$ & $\varphi_1$ & $\theta_2$ & $\varphi_2$ \\
		\midrule
		D3 & $\bullet$ & $\bullet$ & $\bullet$ & $\bullet$ &  & & & & & \\
		D$5$ & $\bullet$ & $\bullet$ & $\bullet$ & & $\bullet$ & $\psi(r)$ & $\bullet$ & $\bullet$ & & \\
		\bottomrule
	\end{tabular}
	\caption{\small Choice of the D5-brane embedding.}
	\label{tab:D5coords}
\end{table}

The dynamics of the D5-branes in the probe approximation regime is governed by the 
DBI action
\begin{equation}
\label{eq:DBI}
S=-T_{\rm D5} N_5\int d^6 \sigma  \sqrt{-\det (\gamma_{ab})}\, ,
\end{equation}
where $T_{\rm D5}=\left((2\pi)^{5}g_s\alpha'{}^3\right)^{-1}$ is the D5-brane tension and $\gamma_{ab}$ is  given by 
\begin{equation}
\gamma_{ab}\equiv G_{\mu\nu} \partial_a X^\mu \partial_b X^\nu+2\pi \alpha' F_{ab} = g_{ab}+2\pi \alpha' F_{ab}\,, 
\end{equation}
$g_{ab}$ being the induced metric on the D5-brane worldvolume and $F=d A$ being the field strength of the U(1) gauge field $A$ living on the D5.
Note that we do not include the Wess-Zumino term in the action since it will not play any role in our setup. 

With our ansatz for the embedding the induced geometry of the D5-brane turns out to be
\begin{equation}\label{eq:wvmetric}
ds^2=g_{ab}\, d\sigma^a d\sigma^b=r^2\left(-dt^2+dx^2+dy^2\right)+\frac{dr^2}{r^2}\left(1+(r\psi')^2\right)+\sin^2\psi \, d^2\Omega_2\, .
\end{equation}

In order to have a finite charge density, an external magnetic field orthogonal to the defect and a longitudinal electric field  
we make the following choice for the worldvolume gauge field (in the $A_r=0$ gauge)
\begin{equation}\label{eq:wvgauge}
2 \pi \alpha' A = A_t(r) dt + \left(E\, t+A_x(r)\right) dx +  \left(B\, x+A_y(r)\right) dy\, .
\end{equation}
The $A_t(r)$ term is the one responsible for the finite charge density, $E$ and $B$ are constant background electric and magnetic fields 
along the $x$ and $z$ directions respectively.\footnote{Note that we included the factor $2\pi \alpha'$ inside all the $A_a$ components and 
	consequently also inside the electric and magnetic fields $E$ and $B$.}
The two functions $A_x(r)$ and $A_y(r)$, as we will shortly see, are in general necessary in order to have a physical configuration; indeed they encode the information about the optical and Hall currents.

Plugging the induced metric \eqref{eq:wvmetric} and the worldvolume gauge field \eqref{eq:wvgauge} into the DBI action \eqref{eq:DBI} and integrating
over all the worldvolume coordinates but $r$ we get
\begin{align}
\label{eq:bgaction}
S= &- \CN_5  \int dr \, \CL\left[\psi(r),\psi'(r),A'_a(r);r\right] \, ,\nonumber \\
\CL\left[\psi(r),\psi'(r),A'_a(r);r\right] = &\frac{\sin^2 \psi(r)}{r^4} \Big[\left(1+\left(B^2-E^2\right)r^4\right) \left(1+r^2 \psi'{}^2(r)\right)  \\
&+  r^4 \left(-A_t'(r)^2+A_y'(r)^2+A_x'(r)^2 -r^4 \left(B A_t'(r)+E A_y'(r)\right)^2\right)\Big]^{1/2}\, , \nonumber
\end{align} 
where $\mathcal{N}_5=4\pi \,T_{\rm D5} N_5 V_{2+1}$,
with $V_{2+1}$ the volume of the (2+1)-dimensional space-time.
We immediately see that $A_t$, $A_x$ and $A_y$ are cyclic coordinates and thus their conjugate momenta, that represent the charge density $\rho$ and the currents $j_x$ and  $j_y$ respectively, are constant.
They turn out to be 
\begin{equation}
\label{eq:momenta}
\rho=\frac{1}{\CN_5}\frac{\partial \CL}{\partial A'_t}\, , \qquad   j_x=\frac{1}{\CN_5} \frac{\partial \CL}{\partial A'_x}\, , \qquad  j_y=\frac{1}{\CN_5} \frac{\partial \CL}{\partial A'_y} \, .
\end{equation}
The presence of cyclic coordinates simplifies considerably the problem since we can immediately solve the 
relations \eqref{eq:momenta} for the gauge field functions $A_a(r)$.
It is also useful to consider the Routhian (density) $\CR$, \ie the Legendre transformed Lagrangian with respect to the cyclic coordinates,
which it is given by
\begin{equation}\label{eq:routhian}
\mathcal{R} = \CN_5 \frac{\mathrm{sgn}(\xi)}{r^4} \sqrt{\left(\xi \chi -a^2\right)\left(1+r^2 \psi '(r)^2\right) }\, ,
\end{equation}
where
\begin{align}
\label{eq:xi}
\xi &=1+\left(B^2-E^2\right) r^4 \, ,\\
\label{eq:chi}
\chi &= \sin^4 \psi(r)+ r^4\left(\rho^2 -j^2_x - j^2_y\right)\, , \\
\label{eq:a}
a&= r^4\left(j_y\, E+ \rho\, B\right)\, .
\end{align}
The equation of motion for the only non trivial variable $\psi$ is then simply given by the Euler-Lagrange equation for the Routhian.

We could think of the conserved momenta $\rho$, $j_x$ and $j_y$ as parameters for the various physical configurations of the system, just like the
external fields $E$ and $B$. However, as we will see in the next subsection, this is only true for the charge density $\rho$, 
since the currents are actually subject to physical constraints that uniquely fix their values in terms of the other parameters.

\subsection{The currents}
\label{subsec:currents}

If we take a careful look at the expression \eqref{eq:routhian} for the Routhian we notice a potentially critical issue. 
The square root term $\sqrt{\xi \chi -a^2}$ seems quite dangerous since it can become imaginary for certain regions of 
the brane worldvolume.  Indeed from Eq.s~\eqref{eq:xi}--\eqref{eq:a} we get that
\begin{equation}\label{eq:xichiminusa2}
\xi \chi -a^2 = \sin^4 \psi +\left[\left(B^2-E^2\right)\sin^4 \psi +\rho^2 -j^2_x - j^2_y\right] r^4 +\left[\left(E^2-B^2\right) j_x^2-\left(E \, \rho +B\, j_y\right)^2\right] r^8 \, . 
\end{equation}
We see that near the boundary $\xi \chi -a^2\simeq \sin^4 \psi$, \ie it is positive, and the Routhian is real. 
However moving toward the Poincar\'e horizon this term may change sign. 
If we want to have a physically acceptable configuration we have to avoid this. Now we examine the conditions that are needed
in order for the Routhian to stay real: we will distinguish two cases, $E>B$ and $E<B$.

\subsubsection*{Currents for $E>B$}

When $E>B$ it is simple to understand what can cause problems to the Routhian. 
Indeed from the definition of $\xi$ in \eqref{eq:xi} we see that in this case there exists a zero of $\xi$ for
a finite positive value of $r=r_s$, given by
\begin{equation}\label{eq:singshell}
r_s = \left(\frac{1}{E^2-B^2}\right)^{1/4}\, .
\end{equation}
The locus of points in the brane worldvolume with $r=r_s$ is usually called \textit{singular shell}.
In general it is quite obvious that when $\xi$ is zero the combination $\xi \chi -a^2$ becomes negative and
this results in an imaginary Routhian. 
However, as pointed out by Karch and O'Bannon in Ref.~\cite{Karch:2007pd}, we can prevent this problem 
by requiring that both $\chi$ and $a$ also have a zero at the same point $r_s$.\footnote{In Ref.~\cite{Karch:2007pd} 
	the procedure was introduced in the study of the D3-D7 system with finite density and external electric field. The procedure was then 
	extended to include magnetic field in \cite{OBannon:2007cex}.}
Imposing this condition fixes the values of the currents $j_x$ and $j_y$ to
the following expressions
\begin{equation}\label{eq:currentsEgtB}
j_x= \frac{\sqrt{\sin^4\psi(r_s) \left(E^2-B^2\right)\left(\rho^2+E^2\right)}}{E} \, , \qquad j_y=-\frac{\rho \,B}{E}\, .
\end{equation}

\subsubsection*{Currents for $E<B$}

When the electric field is smaller that the magnetic field the singular shell coincides with the Poincar\'e horizon.
Nevertheless, also in this case, in order to fix the currents we can look at the sign of $\xi \chi -a^2$ in Eq.~\eqref{eq:xichiminusa2}. 
In particular we observe that this is positive near the boundary while it is negative near the Poincar\'e 
horizon where the $r^8$ contribution dominates.
It is easy to check that in order for $\xi \chi -a^2$ to be always positive we have to choose the currents so as 
to cancel this $r^8$ contribution. In this way we obtain the following values for the currents
\begin{equation}\label{eq:currentsEltB}
j_x=0\, , \qquad j_y =- \frac{\rho\, E}{B}\, .
\end{equation}

\subsection{D5-brane configurations}

In order to build all the possible configurations for the D5-brane embeddings we have to explicitly solve the equation of motion 
for $\psi$ coming from the Routhian \eqref{eq:routhian}. 
We look for solutions that have the following asymptotic behavior near the boundary
\begin{equation}\label{eq:bcpsi}
\psi \simeq \frac{\pi}{2} + c_2 r^2 + \dots\, .
\end{equation}
In principle also a term $c_1 r$ could be present in this expansion but we discard it since $c_1$ would correspond to the 
mass of the fermions in the dual defect theory and in real graphene this is zero. 
The modulus $c_2$ is instead proportional to the chiral condensate, $c_2\sim \langle\bar{f}f\rangle$.

Setting $c_2=0$ gives the trivial constant solution $\psi=\pi/2$. This solution corresponds to the chirally symmetric 
configuration, which we denote $\chi^{S}$. 
Solutions with $c_2\neq 0$ represent instead configurations with spontaneously broken chiral symmetry, $\chi^{SB}$.

The solutions can be classified in black hole (BH) embeddings and 
Minkowski (Mink) embeddings, according to whether or not the brane worldvolume reaches the Poincar\'e horizon \cite{Mateos:2006nu,Kobayashi:2006sb}.
Minkowski embeddings are those for which the worldvolume pinches off at some finite radius $r_0$, \ie $\psi(r_0)=0$.
For such particular configurations the arguments of the previous subsection do not apply. Indeed in this case 
the singular shell does not actually exist, since $r_s>r_0$. 
Thus the on-shell Routhian is always real and we do not need to impose any physical condition on the currents; 
in this case the currents can be safely set to zero. 
Table~\ref{tab:currents} summarizes the values of the currents for all the possible D5-brane embeddings.
\begin{table}[!htb]
	\centering
	\begin{tabular}[c]{p{1cm}ccc} 
		\toprule
		& \multirow{2}{*}{Mink embeddings} &  \multicolumn{2}{c}{BH embeddings}\\
		\cmidrule(r){3-4}
		& & $E>B$ & $E<B$\\
		\midrule
		$j_x$  & 0 & $\dfrac{\sin\psi^2(r_s)\sqrt{\left(E^2-B^2\right)\left(\rho^2+E^2\right)}}{E} $ &  0   \\
		\addlinespace[3mm]
		$j_y$ & 0 & $- \dfrac{\rho\, B}{E}$  & $- \dfrac{\rho\, E}{B}$  \\
		\bottomrule
	\end{tabular}
	\caption{Currents for all the possible D5-brane configurations.}
	\label{tab:currents} 
\end{table}

Note that Minkowski embeddings are possible only for neutral configurations, $\rho=0$ 	\cite{Mateos:2006nu,Kobayashi:2006sb}. 
This is due to the fact that in the string 
picture charge carriers are represented by F1-strings stretching from the D3-branes to the D5-branes. 
Since F1-strings have greater tension than D5-branes they eventually pull the D5-worldvolume 
to the Poincar\'e horizon giving rise to BH embeddings.

\subsection{Phase diagram}

In order to derive the phase diagram for the system we have to compare the free energies of all the possible solutions
in some thermodynamical ensamble, in order to determine which configuration is energetically favored. 
We choose to work in the ensamble where the density $\rho$, the magnetic field $B$ and 
the electric field $E$ are kept fixed. With this choice the right quantity that defines the free energy is 
the on-shell Routhian. 

In the explicit computations of the solutions and their free energies it is actually convenient to reduce the 
number of relevant parameters (\ie the dimension of the phase space) from three to two.  
This can be done, without loss of generality, thanks to the underlying conformal symmetry of the theory.
We choose to measure everything ($\rho$ and $E$ for instance) in units of magnetic field $B$. 

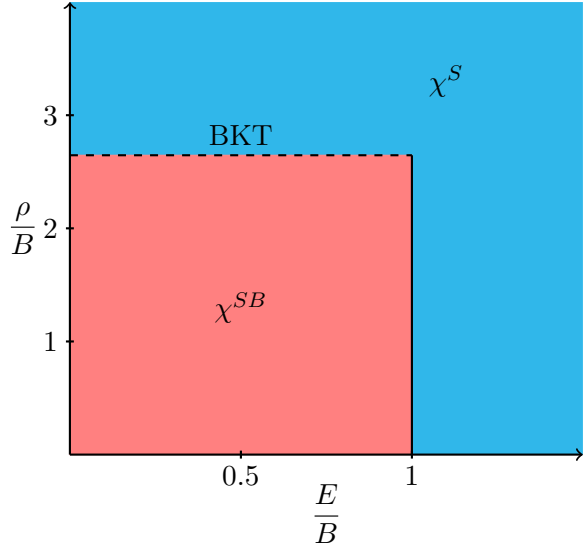
\begin{figure}[!tb]
	\centering
	\begin{tikzpicture}[xscale=4.5,yscale=1.5]
	\fill [Cerulean!70] (0,2.646) -- (0,4) -- ++(1.5,0) -- (1.5,0) 
	-- (1,0) -- ++(0,2.646) -- cycle;
	\filldraw[red!50] (0,0) -- ++(0,2.646) -- ++(1,0) -- (1,0) -- cycle;
	
	\draw [thick, dashed] (0,2.646) -- ++(1,0) node[midway, above] {BKT};
	\draw [thick] (1,0) -- ++(0,2.646);
	
	\draw [thick] [->] (0,0)--(1.5,0) node[midway, below=7pt] {$\dfrac{E}{B}$};
	\foreach \x in {0.5,1}
	\draw[xshift=\x cm, thick] (0pt,-1pt)--(0pt,1pt) node[below=2pt] {$\x$};
	
	\draw [thick] [->] (0,0)--(0,4) node[midway, left=8pt] {$\dfrac{\rho}{B}$};
	\foreach \y in {1,...,3}
	\draw[yshift=\y cm, thick] (-0.33pt,0pt)--(0.33pt,0pt) node[left=2pt] {$\y$};
	
	\node at (0.5,1.3)  {$\chi^{SB}$};
	\node at (1.1,3.3)  {$\chi^{S}$};
	
	\end{tikzpicture}
	\caption{Phase diagram for the D3/D5 system: the blue region covers the chirally symmetric phase, $\chi^{S}$, 
		and the red region the spontaneously broken phase, $\chi^{SB}$.}
	\label{fig:phasediag}
\end{figure}

The results of the analysis on the thermodynamics of the phases can be found in \cite{Evans:2011tk}.\footnote{In this paper 
	the authors consider the same D3/D5 system but also at finite temperature.} 
They are summarized by the phase diagram in Figure~\ref{fig:phasediag}.  
The two competing phases are the chirally symmetric one $\chi^{S}$ (blue region)  and the chirally broken one $\chi^{SB}$ (red region).
Analyzing the phase diagram through a vertical slicing we see that  
when $E<B$, while increasing $\rho$, the system undergoes a BKT transition at $\rho/B=\sqrt{7}$ from
the $\chi^{SB}$ phase to the $\chi^{S}$ one. 
When $E>B$ instead only the trivial $\psi=\pi/2$ solution is allowed and thus 
the system is always in the symmetric phase $\chi^{S}$. 
In the non-symmetric region we have also to distinguish the zero density slice from the finite density area, since in the former 
the D5-brane configurations are  Minkowski embeddings while in the latter BH embeddings.

\section{The fluctuations}
\label{sec:fluctuations}

In this section we review how to introduce the fluctuations for the D3/D5 system and we show their equations of motion. 
We will do this by deriving the effective action for the fluctuation fields \cite{Mas:2008jz,Kim:2011qh}.
At first, the effective action and its equations of motion will be constructed for a 
generic setup of the D3/D5 system and we will eventually specialize it to the case of interest.

\subsection{The effective action for the fluctuations and the open string metric}

As we discussed in the previous section, in the low energy limit, the dynamic of the D3/D5 system 
is encoded in the DBI action showed in Eq.~\eqref{eq:DBI}. 
We use the static gauge where the embedding functions $X^{\mu}$ are split in the following two groups  
\begin{equation}
\begin{cases}
X^a=\sigma^a &\qquad a=0,1,\dots,5 \\
X^I=Z^I (\sigma) & \qquad I=6,\dots,9
\end{cases}
\end{equation}
Exploiting the absence of mixed terms $G_{aI}$ in the background $\ads_5\times S^5$ metric \eqref{eq:ads5}, we can simply write  
the pull-back metric tensor $g_{ab}$ as
\begin{equation}
\label{eq:pullbackmetric}
g_{ab}=G_{ab}(\sigma,Z(\sigma))+G_{IJ}(\sigma,Z(\sigma))\frac{\partial Z^I}{\partial\sigma^a}\frac{\partial Z^J}{\partial\sigma^b}\, .
\end{equation} 
The embedding functions $Z^I$ and the gauge fields $A_a$ can be written as sums of background terms and small perturbations
\begin{align}
\label{eq:fieldsexp}
Z^I(\sigma) & = Z^{(0)I} (\sigma) + \epsilon Z^{(1)I} (\sigma) \, , \nonumber \\
A_a(\sigma)& =A_{a}^{(0)}(\sigma) +\epsilon A_{a}^{(1)}(\sigma) \, ,
\end{align} 
where $\epsilon$ is just a small constant parameter controlling the perturbative expansion. 
The background functions $ Z^{(0)I}$ and $A_{a}^{(0)}$ represent the profile and the gauge potential of the branes
determined by the DBI equations of motion. $Z^{(1)I}$ and $A_{a}^{(1)}$ encode the fluctuations around that background. 

The strategy to build the effective action for the fluctuations is to expand the Lagrangian density up to the second order in $\epsilon$
\begin{equation}
\label{eq:Lagexp}
\mathcal{L}=\mathcal{L}_0 +\epsilon\mathcal{L}_1 +\epsilon^2 \mathcal{L}_2\, .
\end{equation}
We start by considering the expansion of the pull-back metric \eqref{eq:pullbackmetric} 
and the field strength $F_{ab}=\partial_a A_b -\partial_b A_a$
\begin{equation}
g_{ab}=g_{ab}^{(0)}+\epsilon g_{ab}^{(1)}+\epsilon^2 g_{ab}^{(2)}, \quad \quad F_{ab}=F_{ab}^{(0)}+\epsilon F_{ab}^{(1)}\, ,
\end{equation}
where
\begin{equation}
\label{eq:metric_comp}
\begin{split}
& g_{ab}^{(0)}=G_{ab}(\sigma,Z^I (\sigma))\, , \\ 
& g_{ab}^{(1)}=\bigl(G_{ab,K}+G_{IJ,K} Z_{,a}^{(0)I} Z_{,b}^{(0)J}\bigr) Z^{(1)K}+2G_{IJ}Z_{,(a}^{(0)I} Z_{,b)}^{(1)J}\, , \\ 
& g_{ab}^{(2)}=\frac{1}{2}\bigl(G_{ab,KL}+G_{IJ,KL} Z_{,a}^{(0)I} Z_{,b}^{(0)J}\bigr)Z^{(1)K}Z^{(1)L}+G_{IJ}Z_{,a}^{(1)I}Z_{,b}^{(1)J}+2G_{IJ}Z_{,(a}^{(0)I} Z_{,b)}^{(1)J}Z^{(1)K}\, .
\end{split}
\end{equation}
In this way we obtain that the terms in the expansion \eqref{eq:Lagexp} of the Lagrangian are given by\footnote{The procedure and the notation
	used follow faithfully Appendix A of paper~\cite{Mas:2008jz}.}
\begin{equation} 
\label{eq:Lagrangians}
\begin{split}
& \mathcal{L}_0 =\sqrt{\gamma} \, ,\\
& \mathcal{L}_1 =\sqrt{\gamma}\Bigl(\frac{1}{2} \Tr \Sigma^{(1)}\Bigr) \, , \\
& \mathcal{L}_2 = \sqrt{\gamma}\Bigl(\frac{1}{2} \Tr \Sigma^{(2)}-\frac{1}{4} \Tr \Sigma^{(1)2}+\frac{1}{8}(\Tr \Sigma^{(1)})^2 \Bigr) \, ,
\end{split}
\end{equation}
with
\begin{equation}
\label{Gstrana}
\Sigma^{(i)a}{}_b = \gamma^{ac}{\mathcal{G}^{(i)}}_{cb}\, ,
\end{equation}
and
\begin{equation}
\gamma_{ab} = \mathcal{G}_{ab}^{(0)}=g_{ab}^{(0)}+F_{ab}^{(0)}, \qquad \mathcal{G}_{ab}^{(1)}=g_{ab}^{(1)}+F_{ab}^{(1)}, \qquad \mathcal{G}_{ab}^{(2)}=g_{ab}^{(2)}\, .
\end{equation}

Clearly, in order to obtain the effective action for the fluctuations, the quantity we are interested in is just $\CL_2$.\footnote{$\mathcal{L}_0$ is just the DBI Lagrangian for the background fields, which we already studied in the previous section; $\mathcal{L}_1$ vanishes when evaluated on 
	the background profiles.}
Now we want to express this Lagrangian in terms of the so-called \emph{open string metric}, $s_{ab}$, which represents 
the effective geometry seen by open strings in the presence of external fields \cite{Seiberg:1999vs,Gibbons:2000xe}.  
The inverse open string metric $s^{ab}$ ($s_{ab}s^{bc}=\delta_a^c$) can be defined as the symmetric part of the inverse $\gamma_{ab}$ matrix
\begin{equation}
\label{eq:osm_theta}
\gamma^{ab}=(\gamma_{ab})^{-1}=s^{ab}+\theta^{ab} \, ,
\end{equation}
with $s^{ab} = s^{ba}$ and $\theta^{ab} = -\theta^{ba}$. 
This equation can be inverted and it can be shown that
\begin{equation}
\label{eq:OSM}
s_{ab}=g_{ab}-(Fg^{-1}F)_{ab}
\end{equation}
provides the definition of the open string metric as a combination of the pull-back metric and the gauge fields.

With our choice for the D5-brane embedding (see Table~\ref{tab:D5coords}) the worldvolume coordinates are
$\sigma^a=(t,x,y,r,\theta_1,\varphi_1)$. However we will not consider fluctuations along the $S^2$ wrapped by the D5-branes.
This means in particular that $A^{(1)}{}_{\theta_1}=A^{(1)}{}_{\varphi_1}=0$ thus 
the indices in the gauge field fluctuations effectively vary only along $(t,x,y,r)$.
Given that and using Eq.~\eqref{Gstrana} and Eq.~\eqref{eq:OSM} we can write the effective action $S_{\rm eff}\sim \int \CL_2$ as%
\footnote{This effective action is analogous to the one 
	showed in \cite{Kim:2011qh} for D7-probe-branes; however it is more general since it also includes 
	the scalar fluctuations $Z^{(1)I}$ besides the gauge ones.}
\begin{align} 
\label{eq:eff_action}
S_{\rm eff}=-\CN_5 &\int d^6\sigma  \Biggl\{\frac{\sqrt{s}}{g_{5}^2}\Biggl[  \frac{1}{2}s^{ab}g^{(2)}_{ba} 
-\frac{1}{4} s^{ac}s^{bd}g^{(1)}_{cb}g^{(1)}_{da}+\frac{1}{8}\Bigl(s^{ab}g^{(1)}_{ab} \Bigr)^2
-\frac{1}{4}\theta^{ac}\theta^{bd}g^{(1)}_{cb}g^{(1)}_{da} \nonumber \\
&+ \frac{1}{4}s^{ac}s^{bd}F^{(1)}_{cb}F^{(1)}_{da}
-\frac{1}{2} (s^{ac}\theta^{bd}+\theta^{ac}s^{bd})g^{(1)}_{cb}F^{(1)}_{da}   
- \frac{1}{4}s^{ab}\theta^{cd}{g^{(1)}}_{ab} F^{(1)}_{cd}\Biggr] \\
&+\frac{1}{8}Q\epsilon^{abcd\theta_1\varphi_1}F^{(1)}_{ab}F^{(1)}_{cd} \Biggr\}\, , \nonumber
\end{align}
with $s= - \det s_{ab}$, $g_5^2 = \dfrac{\sqrt{s}}{\sqrt{\gamma}}$ and 
\begin{equation} 
\label{eq:topQ}
Q = -\frac{\sqrt\gamma}{8}\epsilon_{efgh\theta_1\varphi_1}\theta^{ef}\theta^{gh}\, ,
\end{equation}
where the Levi-Civita symbol is defined as $\epsilon^{txyr\theta_1\varphi_1}=-\epsilon_{txyr\theta_1\varphi_1}=1$. 
Note that the last term of $S_{\rm eff}$ is a topological term that appears only if 
there are two non-vanishing components of $\theta^{ab}$ with all different indices in the subset $a,b=(t,x,y,r)$.

We can now plug the pull-back metric components \eqref{eq:metric_comp} into the effective action in order to write it
as a sum of kinetic terms, mass terms and interaction terms for the fluctuating scalar fields and gauge fields
\begin{align}
\label{eq:eff_action2}
S_{\rm eff}=-\CN_5 &\int d^6\sigma  \Biggl\{ \frac{\sqrt{s}}{g_{5}^2}\Bigl[ \mathcal{P}^{ab}_{IJ}\, Z^{(1)I}_{,a}Z^{(1)J}_{,b}+\mathcal{Q}_{IJ}\, Z^{(1)I}Z^{(1)J}+\mathcal{R}_{IJ}^{a} \, Z^{(1)I}_{,a}Z^{(1)J}+\frac{1}{4}s^{ac}s^{bd}F^{(1)}_{cb}F^{(1)}_{da} \nonumber \\
+& \mathcal{S}_{I}^{abc}Z^{(1)I}_{,a}F^{(1)}_{bc}+\mathcal{T}_{I}^{ab}\, Z^{(1)I}F^{(1)}_{ab}\Bigr]+
\frac{1}{8}Q\epsilon^{abcd\theta_1\varphi_1}F^{(1)}_{ab}F^{(1)}_{cd} \Biggr\} \, ,
\end{align}
where the coefficients $\mathcal{P}, \mathcal{Q}, \mathcal{R}, \mathcal{S}, \mathcal{T}$ are
\begin{align}
\label{eq:coefficients}
\mathcal{P}_{IJ}^{ab}\equiv & \frac{1}{2}\Bigl(G_{IJ}s^{ab}-G_{IK}G_{JL} Z^{(0)K}_{,c} Z^{(0)L}_{,d}(s^{ad}s^{bc}-s^{ac}s^{bd}+s^{ab}s^{cd}+\theta^{ad}\theta^{bc}-\theta^{ab}\theta^{cd})\Bigr)\, , \nonumber \\
\mathcal{Q}_{IJ}\equiv & \frac{1}{8}\Bigl[2s^{ab}G_{ab,IJ}+2s^{ab}G_{KL,IJ}Z^{(0)K}_{,a}Z^{(0)L}_{,b}-(2s^{ac}s^{bd}-s^{ab}s^{cd}-2\theta^{ac}\theta^{bd})\nonumber \\
& \times [G_{ab,I}G_{cd,J}+Z^{(0)K}_{,a} Z^{(0)L}_{,b}(G_{cd,J}G_{KL,I}+G_{cd,I}G_{KL,J}+G_{KL,I}G_{MN,J}Z^{(0)M}_{,c} Z^{(0)N}_{,d})] \Bigr]\, ,\nonumber \\
\mathcal{R}_{IJ}^{a}\equiv & s^{ab}G_{IK,J}Z^{(0)K}_{,b}+\frac{1}{2}G_{cd,J}G_{IJ}Z^{(0)K}_{,b}(s^{ab}s^{cd}-2s^{ac}s^{bd}+2\theta^{ac}\theta^{bd})  \\
& + G_{KL,J}G_{IM}Z^{(0)K}_{,b}Z^{(0)L}_{,c}Z^{(0)M}_{,d}(s^{ad}s^{bc}-2s^{ab}s^{cd}-2\theta^{ab}\theta^{cd})\, , \nonumber  \\
\mathcal{S}_{I}^{abc}\equiv & -\frac{1}{2}G_{IJ}Z^{(0)J}_{,d}(s^{cd}\theta^{ab}-s^{bd}\theta^{ac}+s^{ad}\theta^{bc}-s^{ac}\theta^{bd}+s^{ab}\theta^{cd})
\, , \nonumber \\
\mathcal{T}_{I}^{ab}\equiv & -\frac{1}{4}(G_{cd,I}+G_{JK,I}Z^{(0)J}_{,c}Z^{(0)K}_{,d})(s^{cd}\theta^{ab}-2s^{bc}\theta^{ad}+2s^{ac}\theta^{bd}) \, . \nonumber
\end{align}
%


From the Lagrangian \eqref{eq:eff_action2} we obtain the general equations of motion for both the embedding functions and the gauge fields
\begin{align}
\text{EoM}[Z^{(1)I}] \rightarrow \,
\frac{\sqrt{s}}{g_5 ^2}\Bigl( &2 \mathcal{Q}_{IJ}Z^{(1)J}+\mathcal{R}_{JI}^a Z^{(1)J}_{,a}+\mathcal{T}_{I}^{ab} F^{(1)}_{ab}\Bigr)  \nonumber \\ 
\label{eomPsi}
& -\partial_{a}\Bigl[\frac{\sqrt{s}}{g_5 ^2}\Bigl(2 \mathcal{P}_{IJ}^{ab} Z^{(1)J}_{,b}+\mathcal{R}_{IJ}^a Z^{(1)J}+\mathcal{S}_{I}^{abc} F^{(1)}_{bc} \Bigr) \Bigr]=0 \, , \\
\label{eomGauge}
\text{EoM}[A_b^{(1)}] \rightarrow \, 	\partial_a \Bigl[\frac{\sqrt{s}}{g_{5}^2} & \Bigl(F^{(1)ab} +2 \mathcal{S}_{I}^{cab} Z^{(1)I}_{,c}+2 \mathcal{T}_{I}^{ab}Z^{(1)I} \Bigr)+\frac{Q}{2}\epsilon^{abcd\theta_1\phi_1} F^{(1)}_{cd} \Bigr]=0 \, . 
\end{align}

The coefficients \eqref{eq:coefficients} are very complicated in general, however when we specialize them to the case under consideration 
many simplifications are possible. 
First of all, since we consider background solutions with a non trivial transverse profile of the D5-branes only along 
$\psi$, we will also consider only one scalar perturbation field along the same direction, \ie $Z^{(1) I}=\psi^{(1)}\delta_{I\psi}$.
With this assumption, and using the background specification of Section~\ref{sec:model}, we obtain that the non vanishing components of the coefficients \eqref{eq:coefficients} are just
\begin{equation}\label{eq:coeff_eff_act}
\begin{split}
\mathcal{P}_{\psi\psi}^{ab}=&\frac{1}{2}\Bigl[s^{ab}(1-s^{rr}\psi'^2 ) -\psi'^2 \theta^{ar}\theta^{br}\Bigr]\, ,  \\
\mathcal{Q}_{\psi \psi}=&\frac{\cos 2\psi}{\sin^2\psi}\, , \\
\mathcal{R}_{\psi \psi}^a=& \, 2\cot\psi \, \psi' \, s^{ar}\, , \\
\mathcal{S}_{\psi}^{abc}=&-\frac{1}{2} \psi'\bigl(s^{cr}\theta^{ab}-s^{br}\theta^{ac}+s^{ar}\theta^{bc}-s^{ac}\theta^{br}+s^{ab}\theta^{cr}\bigr)\, ,  \\
\mathcal{T}_{\psi}^{ab}=&-\cot\psi \, \theta^{ab} \, .
\end{split}
\end{equation}
To simplify the notation we denote the background profile $\psi^{(0)}$ simply as $\psi$.  

\section{The conductivities}
\label{sec:conductivities}

In this section we show the results for the Ohm and Hall conductivities obtained from 
the holographic D3/probe D5 model introduced in Section~\ref{sec:model}.
Notice that the DC conductivities are already known, since by definition 
they can be simply calculated from the currents $j_x$ and $j_y$:
\begin{equation}\label{eq:def_cond}
  j_i = \sigma_{ij} E_j\, .
\end{equation}
Actually, using the currents determined in Section~\ref{subsec:currents}, what we obtain is the full
non-linear DC conductivity tensor. 
In this Section we will instead focus on the linear response theory, that allows us to derive the frequency dependent
conductivities.
As a first step we solve the equation for the gauge field fluctuations $A^{(1)}_a$ with
the following (zero momentum) ansatz 
\begin{equation}\label{eq:ansatz_fluct}
 A^{(1)}_a (t,r)= e^{-i\omega t} a_{a}(r)\, .
\end{equation}
We also fix the gauge choosing $a_r =0$.
Then the conductivities $\sigma_{ij}$ are obtained through the Kubo formula
\begin{equation}
\sigma_{ij}(\omega)= - i \frac{G_{j_i j_j}^R(\omega)}{\omega}\, ,
\end{equation}
where $G_{j_i j_j}^R$ is the retarded current-current Green's function.
Using the holographic dictionary this can be computed as
\begin{equation}\label{eq:hol_cond}
\sigma_{ij}=\lim\limits_{r\rightarrow 0} -\frac{i}{\omega}\, \frac{a'_{i}(r)}{a_j(r)}\, .
\end{equation}

According to what we saw in Section~\ref{sec:model} for the currents, we distinguish 
two cases, $E>B$ and $E<B$.

\subsection{Conductivities for $E>B$}

As showed in the phase diagram of Figure~\ref{fig:phasediag}, when $E>B$ there is only one stable configuration for every 
value of the charge density, namely the chirally symmetric one, with $\psi=\frac{\pi}{2}$.

The coefficients \eqref{eq:coeff_eff_act} of the effective action for the fluctuations become 
extremely simple in this case
\begin{equation}
\label{eq:coeff_sym}
\mathcal{P}_{\psi\psi}^{ab}=\frac{s^{ab}}{2} \, , \qquad
\mathcal{Q}_{\psi \psi}=-1 \, , \qquad
\mathcal{R}_{\psi \psi}^a=0 \, ,\qquad
\mathcal{S}_{\psi}^{abc}=0 \, , \qquad
\mathcal{T}_{\psi}^{ab}=0 \, .
\end{equation}
This basically means that the gauge and scalar fluctuations are decoupled.
Since we are interested in the conductivity we can neglect the scalar field $\psi^{(1)}$. 

From Eq.s~\eqref{eq:osm_theta}-\eqref{eq:OSM} we can compute the open string metric $s_{ab}$ as
\begin{equation} \label{eq:OSMcase1}
\begin{split}
	s_{ab} \, d\sigma^a \, d\sigma^b &= -\frac{1}{r^2}\left(1-\frac{r^4}{r_\rho^4}\right)\, dt^2 
	+ 2\frac{\rho}{E}\frac{r^2}{r_s^2 r_\rho^2}\, dt \, dx 
	-2\frac{B}{E}\frac{r^2}{r_\rho^4}\,  dt\, dy +\frac{r^2}{r_s^2 r_\rho^2} \, dt\, dr \\
	&+\frac{1}{r^2}\left(1+\frac{\rho^2}{E^2}\frac{r^4}{r_\rho^4}\right)\, dx^2    
	-2\frac{\rho B}{E^2}\frac{r^2}{r_s^2 r_\rho^2}\,dx \, dy 
	+2\frac{\rho}{E}\frac{r^2}{r_s^4}\, dx \, dr  \\
	&+\frac{1}{r^2}\left(1+\frac{B^2}{E^2}\frac{r^4}{r_\rho^4}\right)\, dy^2 
	-2\frac{B}{E}\frac{r^2}{r_s^2 r_\rho^2}\, dy\, dr
	+\frac{1}{r^2}\left(1+\frac{r^4}{r_\rho^4}\right) \, dr^2 + d \Omega_1^2 \, ,
\end{split}
\end{equation}
and the antisymmetric tensor $\theta^{ab}$, whose non vanishing components are
\begin{equation}
\label{eq:theta_case1}
\begin{split}
	\theta^{tx}&= E \, r^4 \, , \qquad \theta^{tr}=- \rho \, r^4 \, , \qquad \theta^{xy}=- B \, r^4\,,\\
	&\theta^{xr}=\frac{r^4}{E\, r_s^2 r_\rho^2}\, ,\qquad \theta^{yr}=-\frac{\rho B \,r^4}{E}\, ,
\end{split}
\end{equation}
where $r_s$ is the singular shell radius introduced in Eq.~\eqref{eq:singshell} and $r_\rho$ is defined as
\begin{equation} 
	r_\rho \equiv \left(E^2+\rho^2\right)^{-1/4}\, .
\end{equation}

Notice that the open string metric \eqref{eq:OSMcase1} is a black hole metric and its horizon radius exactly
coincides with the singular shell radius \eqref{eq:singshell}. 
The Hawking temperature of this black hole geometry is given by
\begin{equation}
\label{eq:Teff}
	T_{\rm eff}=\frac{r_\rho^2}{\pi\, r_s^3\sqrt{B}}=\frac{(E^2-B^2)^{3/4}}{\pi \sqrt{B(\rho^2+E^2)}}.
\end{equation}
and it represents the effective temperature felt by open strings.
Thus, even though we considered a zero temperature background, the presence of the electric field 
induces an effective thermal heat bath. 

Although the theta tensor \eqref{eq:theta_case1} has apparently enough non zero components to give rise to a topological term,
it turns out that when these components are plugged into \eqref{eq:topQ} they yield $Q=0$. Thus the effective action 
for the fluctuations is just given by the Maxwell action.
Nevertheless, due to the form of the open string metric \eqref{eq:OSMcase1}, without vanishing components
in the 4-dimensional $(t,x,y,r)$ sub-manifold (unless for zero density), 
the equations of motion for the gauge fluctuations are still quite complicated.
We can simplify them slightly by making a change of coordinates that kills the mixed 
radial components $s_{tr}$, $s_{xr}$, $s_{yr}$ of the open string metric, in such a way that
the latter becomes\footnote{The change of coordinates is of the form
\begin{equation*}
	dt \to dt + f_{tr}(r) \, dr\,, \qquad dx \to dx + f_{xr}(r) \, dr\,, \qquad dy \to dy + f_{yr}(r) \, dr\, ,
\end{equation*}
with the three $f$ functions fixed in such a way to get rid of the mixed radial components of the open string metric.
It turns out that this change of coordinates does not affect the computation of the conductivities since the behavior of the
$f$ functions near the boundary is of order $O(r^4)$. Thus we can safely proceed with this transformed metric.}
\begin{equation} \label{eq:newOSMcase1}
\begin{split}
s_{ab} \, d\sigma^a \, d\sigma^b &= -\frac{1}{r^2}\left(1-\frac{r^4}{r_\rho^4}\right)\, dt^2 
+ \frac{\rho}{E}\frac{r^2}{r_s^2 r_\rho^2}\, dt \, dx 
+\frac{1}{r^2}\left(1+\frac{\rho^2}{E^2}\frac{r^4}{r_\rho^4}\right)\, dx^2 \\   
&-\frac{\rho B}{E^2}\frac{r^2}{r_s^2 r_\rho^2}\,dx \, dy  
+\frac{1}{r^2}\left(1+\frac{B^2}{E^2}\frac{r^4}{r_\rho^4}\right)\, dy^2 
+\left[r^2\left(1-\frac{r^4}{r_s^4}\right)\right]^{-1} \, dr^2 + d \Omega_1^2 \, .
\end{split}
\end{equation}

Now we have all the ingredients to write down the equations of motion for the gauge fields fluctuations $A^{(1)}_a$
using the ansatz \eqref{eq:ansatz_fluct} and the gauge choice $a_r=0$.
The $a_t$ component can be easily decoupled and 
one is left just with the equations of motion for $a_x$ and $a_y$.
%
%
%
%
In the near-horizon limit, $r\to r_s$, both these equations become
\begin{equation}
\label{eq:nearshell}
a''_{i}+\frac{1}{r-r_s}\,a'_{i}+\frac{\omega^2\, B\, r_s^6}{16\, r_\rho^4 \, (r-r_s)^2}\,a_{i}=0\, \qquad i=x,y. 
\end{equation}
Performing a Frobenius expansion we get that the correct behavior near the open string metric horizon is
\begin{equation}
\label{eq:shellexp}
a_i(r)_{r\rightarrow\hspace{1pt} r_s}\rightarrow \hspace{2pt} (r-r_s)^{\pm i\frac{\omega}{4\pi T_{\rm eff}}}\, ,
\end{equation}
where $T_{\rm eff}$ is the effective temperature in Eq.~\eqref{eq:Teff}.
Therefore, near the singular shell we can write the solution as  
\begin{equation}
\label{eq:shellbehaviour}
a_i(r)=(r-r_s)^{ i\frac{\omega}{4\pi T_{\rm eff}}} \chi_i(r),
\end{equation}
where the first term takes into account the right infalling behavior near the singular shell while $\chi$ is a regular function which can be expanded analytically in powers of $(r-r_s)$.
In particular we can express the near-singular shell shape of $\chi_i$ as follows
\begin{equation}\label{eq:nearSS}
\chi_x^{h}=c_{x_0}+c_{x_1}(r-r_s)+c_{x_2}(r-r_s)^2+\dots \quad \chi_y^{h}=c_{y_0}+c_{y_1}(r-r_s)+c_{y_2}(r-r_s)^2+\dots
\end{equation}
where the coefficients $c_{x_1},c_{x_2},\dots,c_{y_1},c_{y_2},\dots$ can be easily determined as functions of $\omega,r_s,\rho$ and of the two 
moduli $c_{x_0}$, $c_{y_0}$.

\subsubsection*{DC conductivities}

The DC conductivities can be easily extracted from the equations of motion in an analytical way, since
they only require the knowledge of the solution for the fluctuations up to the linear order in $\omega$
in the small frequency limit.
Then the strategy is to expand the functions $\chi_x$ and $\chi_y$ in powers of the frequency $\omega$ as follows
\begin{equation}
\label{eq:chiDC}
\chi_x(r)=\chi_{x}^{(0)}+\omega \chi_{x}^{(1)}, \qquad \chi_y(r)=\chi_{y}^{(0)}+\omega \chi_{y}^{(1)} \, .
\end{equation}
The solutions for the $\chi_i^{(k)}$ functions can be obtained analytically. 
Imposing regularity at the singular shell and using the holographic formula \eqref{eq:hol_cond} at the leading order in $\omega$
we found the following results for the DC conductivities
\begin{equation}
\sigma_{xx}^{\rm DC}=\frac{E^4+B^2\,\rho^2}{E^2\sqrt{(E^2-B^2)(E^2+ \rho^2)}}, \quad \sigma_{yy}^{\rm DC}= \frac{\sqrt{(E^2-B^2)(E^2+\rho^2)}}{E^2},\quad \sigma_{xy}^{\rm DC}=\sigma_{yx}^{\rm DC}=\frac{\rho \, B}{E^2}.
\end{equation}
It is straightforward to check that these  conductivities are in perfect agreement with the 
expressions of the currents in Eq.~\eqref{eq:currentsEgtB}. 
Indeed we can extract the linear conductivities from the latter as follows:
we add a small perturbation $\varepsilon$ along the $j=x,y$ direction to the background electric field, $\vec{E} \to \vec{E} + \varepsilon \hat{j}$ 
($\hat{j}$ unit vector pointing along $j$) and then, according to  Eq.~\eqref{eq:def_cond},
we read the conductivities $\sigma_{ij}$ as the coefficients of the linear term in $\varepsilon$ of the current $j_i$.

\subsubsection*{AC conductivities}

In order to compute the full frequency dependent conductivity we have to solve the equations for the fluctuations and then plug
the solutions in the formula \eqref{eq:hol_cond}. Though linear, these equations cannot be solved analytically
so we used a numerical technique.\footnote{We simply used the built-in \texttt{NDSolve} command in Mathematica.}
The boundary conditions of the differential equations are fixed at the singular shell using \eqref{eq:shellbehaviour} and \eqref{eq:nearSS}.

In the following we show some plots of the conductivities computed with our model in the $E>B$ sector. 
Without loss of generality, the magnetic field $B$ has been set to $1$ in all plots.

\begin{figure}[htbp]
	\centering
		\includegraphics[width=.48\textwidth]{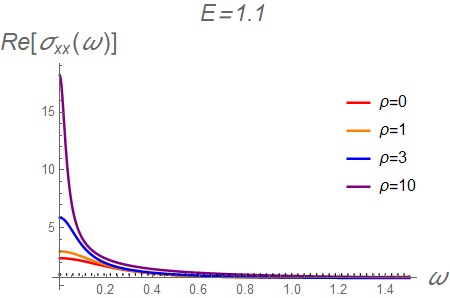}
		\hfill
		\includegraphics[width=.48\textwidth]{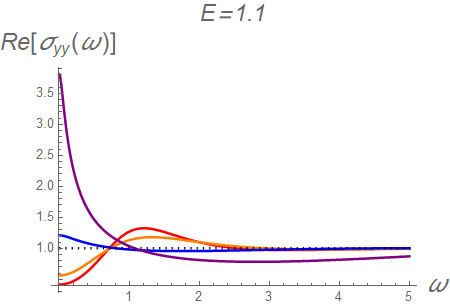} 
		\\
		\includegraphics[width=.48\textwidth]{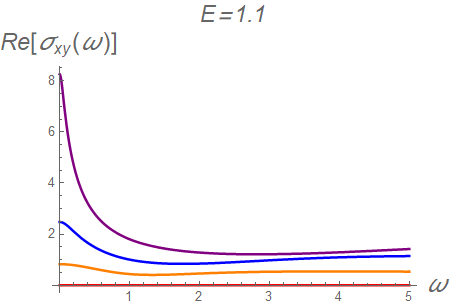}
		\hfill
		\includegraphics[width=.48\textwidth]{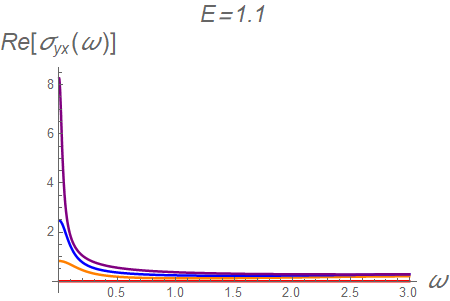}
		\\
		\includegraphics[width=.48\textwidth]{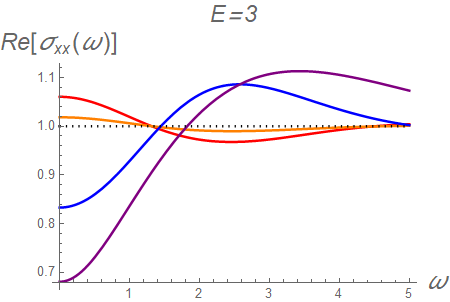}
		\hfill
		\includegraphics[width=.48\textwidth]{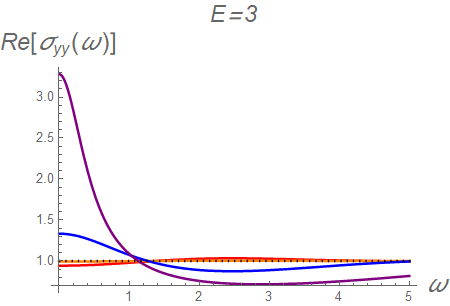}
		\\
		\includegraphics[width=.48\textwidth]{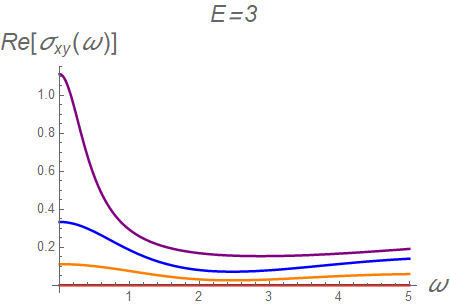}
		\hfill
		\includegraphics[width=.48\textwidth]{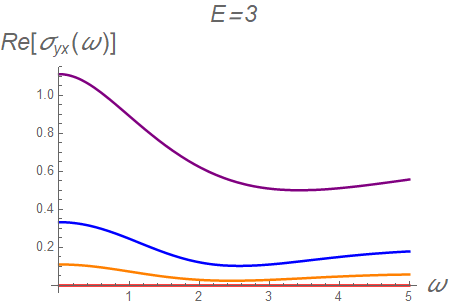}
	\caption{Real parts of the conductivities for different values of electric field $E$ and charge density $\rho$. \label{fig:real_cond_EgtB}  }
\end{figure}

\begin{figure}[!h]
	\centering
	\includegraphics[width=.48\textwidth]{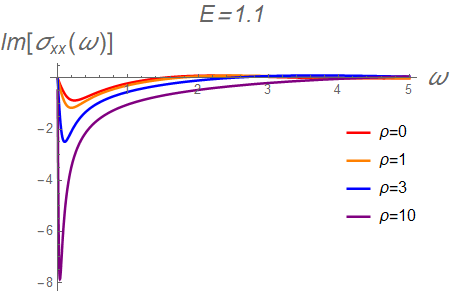}
	\hfill
	\includegraphics[width=.48\textwidth]{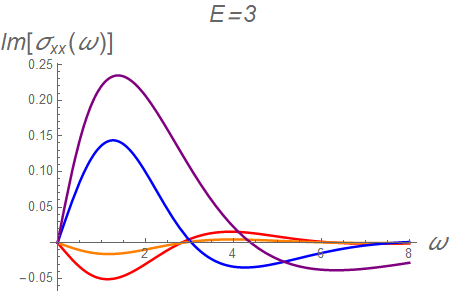}
	\caption{Imaginary part of the longitudinal $\sigma_{xx}$ conductivity for different values of electric field $E$ and charge density $\rho$. \label{fig:Im_cond}}
\end{figure}

Figure~\ref{fig:real_cond_EgtB} shows only the real part of the conductivities, since the imaginary one can be straightforwardly determined by means of the Kramers-Kronig relation, relating the real and the imaginary parts of the retarded Green's function as follows
\begin{equation}\label{eq:kramers-Kronig}
	\imm \left[G_R (\omega) \right]= - \CP \int_{-\infty}^{\infty} \frac{d\, \omega}{\pi} \, \frac{\real \left[G_r(\omega')\right]}{\omega' -\omega}\, ,
\end{equation}
where $\CP$ denotes the principal value of the integral. 
Nevertheless in Figure~\ref{fig:Im_cond} we show some examples of $\imm[\sigma_{xx}(\omega)]$ for completeness.

From Figure~\ref{fig:Im_cond} we observe that the imaginary part of the conductivities goes to zero not only in the high frequency limit, but also in the low frequency one, $\omega \to 0$. This is also true in general for all the other cases.

Looking at the plots in Figure~\ref{fig:real_cond_EgtB} we can immediately note that all the real parts of the conductivities go to a constant in the high frequency limit $\omega \to \infty$, \ie 
\begin{equation}
\lim_{\omega \to \infty} \sigma_{ij} (\omega) = C_{ij}\, .
\end{equation}
This is a standard behavior of the (2+1)-dimensional systems and it is consistent since the conductivities are dimensionless in this case. 

It can be easily checked that the low frequency trend of the conductivities is consistent with the DC values determined by the Karch-O'Bannon method.
Even if the brane system is translationally invariant, we do not see the presence of the delta function Drude peak, 
as it can be argued	 from the imaginary part of the conductivities, which in fact goes to zero in the low frequency limit.\footnote{The
	delta function Drude peak should appear as a pole in the imaginary part of the conductivities at $\omega \to 0$. } 
Furthermore we recall that the analytical property of the sum rule implies that the following relation
\begin{equation}
	\int_{0}^{\infty} {d \omega \,  \left(\real \left[\sigma_{ij} \left(\omega \right) \right] - C_{ij} \right) } = 0
\end{equation}
must hold if there is no delta function peak \cite{Ryu:2011vq}. We checked that this relation is fulfilled for all the longitudinal and transverse conductivities we have considered. 
As it is well known, the reason why we do not see the delta function Drude peak is that we are studying the system in the probe approximation limit
which introduces dissipation without breaking the 
translational invariance: in this regime the gluon sector of the $3+1$ dimensional Super Yang-Mills theory plays the role of the lattice 
in solid state physics and, due to its large density, it can absorb a large amount of momentum standing basically still \cite{Karch:2007pd}. 

Note that as the ratio $E/B \to 1$ the value of the DC conductivity grows and consequently we see the emergence 
of a Drude-like peak at small frequency. This peak eventually becomes a delta function, $\delta(\omega)$,
when $E=B$ and the effective temperature felt by open strings becomes zero. 
This delta function is not related to momentum conservation, but to an additional conserved quantity 
that probe branes have at zero temperature only, the charge current operator \cite{Chen:2017dsy}. 
When the (effective) temperature is small but finite the weak non-conservation of this current is then responsible for 
the appearance in the conductivity of the Drude-like peaks we observed. 
The peaks are indeed associated to poles in the Green's function, whose presence is a general feature in low energy effective theories 
with approximately conserved operators (\emph{quasihydrodynamics}), as pointed out recently in 
\cite{Grozdanov:2018fic}.

Very similar results for the AC conductivity have been found in \cite{Hartnoll:2009ns} considering non vanishing temperature,
charge density and electric field in the context of a neutral Lifshitz-invariant theory and in \cite{Das:2010yw} studying 
rotating D$p$-branes at zero temperature which induce an effective horizon. 
In both of these papers the authors found the same standard behavior at large frequencies and a finite peak in the limit $\omega \to 0$. 
Here we managed to obtain a very similar physics in the context of the D3/probe-D5-brane system. 
This is possible because, although we are considering zero temperature, there is an effective temperature induced by the singular shell, 
which plays the role of an horizon.

\subsection{Conductivities for $E<B$}

When $E<B$ the open string metric takes the form
\begin{equation}\label{eq_osm_EltB}
\begin{split}
	s_{ab}\,d\sigma^a d\sigma^b= &-\frac{\left(r_b^4+r^4\right) \left(B^2 \Delta^4 -E^2 r^4\right)}{B\, r^2 \left(B^2 \Delta^4 r_b^4+r^4\right)}\, dt^2 
	-2\frac{  E \, r^2 \left(r_b^4+r^4\right)}{B^2 \Delta^4 r_b^4+r^4}\, dt\,dy 
	+ \frac{r_b^4+r^4}{B\, r_b^4	r^2}\, dx^2 \\
	&+ \frac{B	\left(r_b^4+r^4\right) \left( \Delta^4+r^4\right)}{r^2 \left(B^2 \Delta^4 r_b^4+r^4\right)}\, dy^2 
	+ \frac{B \,  \Delta^4\left(r_b^4+r^4\right) \left(1+r^2	\psi '(r)^2\right)}{r^2 \left(B^2 \Delta^4 r_b^4+r^4\right)} dr^2
	+ d \Omega_1^2 \, ,
\end{split}
\end{equation} 
where 
\begin{equation}
 	r_b=\left(B^2-E^2\right)^{-1/4}\, , \qquad \Delta = \left(B^2 + \rho^2 \csc(\psi(r))^4\right)^{-1/4} \, .
\end{equation}
The open string metric has no finite radius horizon and 
indeed we know that the singular shell is located at the Poincar\'e horizon. 
The antisymmetric $\theta_{ab}$ tensor is given by
\begin{equation}
\label{eq:theta_EltB}
\begin{split}
\theta^{tx}&= \frac{E\, B\, r_b^4 \, r^4}{r_b^4+r^4} \, , \qquad 
\theta^{tr}=-\frac{\rho\,   r^4 r_b^2}{\Delta ^2 \left(r_b^4+r^4\right) \sin^2\psi(r)}\sqrt{\frac{B^2 \Delta ^4 r_b^4+r^4}{1+r^2 \psi '(r)^2}}\,,\\
\theta^{xy}&=-\frac{ B^2\, r_b^4 \, r^4}{r_b^4+r^4} \, ,\qquad 
\theta^{yr}=-\frac{E\, \rho\,  r^4 r_b^2 }{B\,\Delta ^2 \left(r_b^4+r^4\right) \sin^2\psi(r)}\sqrt{\frac{B^2 \Delta ^4	r_b^4+r^4}{1+r^2 \psi '(r)^2}}\, .
\end{split}
\end{equation}
Differently from the previous case, now the  $\theta_{ab}$ tensor gives rise to a non vanishing topological term, $Q\neq 0$, in the 
effective action for the fluctuations.

In this regime the D3/D5 system has two stable phases,
the chirally symmetric and the chirally broken ones (see Figure~\ref{fig:phasediag}). 
So, we have to further distinguish between these two cases. 

\subsubsection*{Symmetric phase ($\rho > \sqrt{7}\, B $)}

When the charge density $\rho$ is above the threshold value $\sqrt{7}\, B$ the system is still in the chirally symmetric
phase just as for $E>B$. Thus, also in this case, we have that the gauge fluctuations decouple from the scalar ones 
(the coefficients of the effective actions \eqref{eq:eff_action2} are those shown in Eq.~\eqref{eq:coeff_sym}).
So the effective action we need to consider in order to study the gauge fluctuations is given by 
\begin{equation}\label{eq:action_EltB}
S_{\rm eff}=-\CN_5 \int d^6\sigma  \Biggl[\frac{\sqrt{s}}{g_{5}^2}\frac{1}{4}s^{ac}s^{bd}F^{(1)}_{cb}F^{(1)}_{da} +
\frac{1}{8}Q\epsilon^{abcd\theta_1\varphi_1}F^{(1)}_{ab}F^{(1)}_{cd} \Biggr] \, .
\end{equation}
The equations for the $a_x$ and $a_y$ fluctuations in the $r\to \infty$ limit are
\begin{equation}
\label{eq:nearPoincare}
a''_{i}-\frac{2}{r} a'_{i}-\omega^2\, B^3\,r_b^4\,a_{i}=0\, \qquad i=x,y.
\end{equation}
The solutions for the gauge fluctuations near the Poincar\'e horizon, with the desired (infalling) behavior, can 
be written as
\begin{equation}\label{eq:nearPoi_sol}
	a_{i} (r) =  r\, e^{ i \, \omega \, B^{3/2}\, r_b^2\, r} \, \chi_i(r)\, ,
\end{equation}
where again the $\chi_i$ admit a power series expansion near the Poincar\'e horizon. We will use the form of the solutions 
given in Eq.~\eqref{eq:nearPoi_sol} in order to fix the boundary conditions in the numerical integration of the equations of motion.

\subsubsection*{DC conductivities}
Also in this case the zero frequency results for the conductivities can be extracted analytically. They turn out to be
\begin{equation}\label{eq:DC_cond_EltB_1}
\sigma_{xx}^{\rm DC}=0\, , \qquad \sigma_{yy}^{\rm DC}= 0\, , \qquad \sigma_{xy}^{\rm DC}=-\sigma_{yx}^{\rm DC}=\frac{\rho}{B}.
\end{equation}
They are again consistent with the expressions for the currents \eqref{eq:currentsEltB}.

 \subsubsection*{AC conductivities}

 In the following we show some plots of the conductivities for $E<B$ and $\rho>\sqrt{7}\, B$.
 Without loss of generality, the magnetic field $B$ has been set to $1$ in all plots.

 \begin{figure}[!h]
 	\centering
	\includegraphics[width=.48\textwidth]{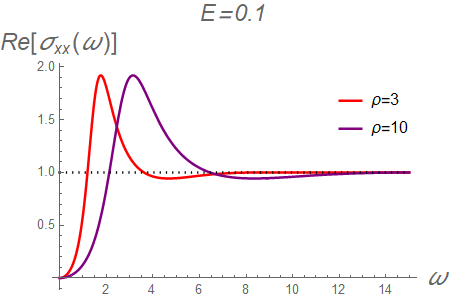}
 	\hfill
 	\includegraphics[width=.48\textwidth]{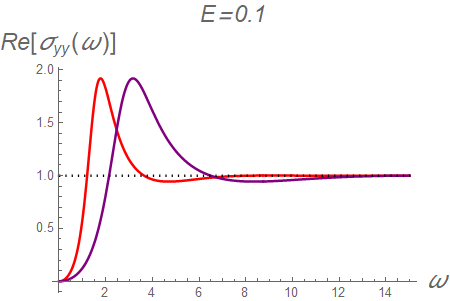}
 	\\
 	\includegraphics[width=.48\textwidth]{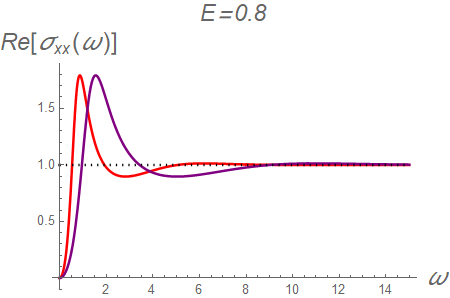}
 	\hfill
 	\includegraphics[width=.48\textwidth]{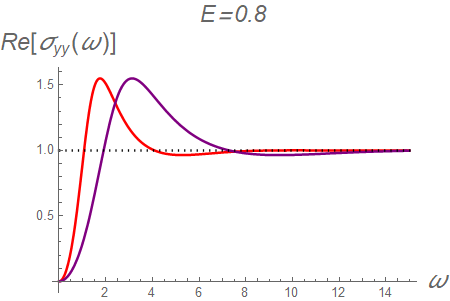}
 	\caption{Real parts of the conductivities for different values of electric field $E$ and charge density $\rho$ in the case $E < B$ (symmetric phase). We do not show the plots of the transverse conductivities since they are just constants in $\omega$. \label{fig:real_cond_EltB}}
 \end{figure}

 Looking at the plots in Figure~\ref{fig:real_cond_EltB} we immediately recognize for the real parts of the longitudinal conductivities the same standard large frequency behavior of the $E > B$ case, \ie we have that they are constant in the limit $\omega \to \infty$. In the low energy limit instead they vanish as they should, in order to be consistent with the DC conductivities. At intermediate frequencies they exhibit a finite peak which becomes larger when the charge density increases. 
 We do not show the plots of the transverse conductivities since they do not depend on $\omega$, 
 so they are identically equal to their DC value, $\sigma_{xy}=-\sigma_{yx}=\frac{\rho}{B}$.
 
\begin{figure}[!h]
 	\centering
 	\includegraphics[width=.48\textwidth]{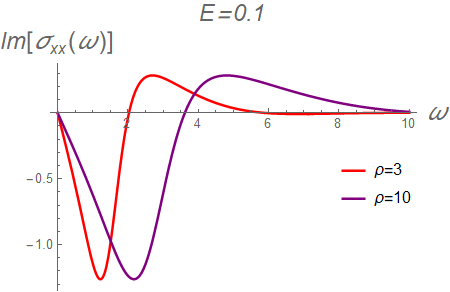}
 	\hfill
 	\includegraphics[width=.48\textwidth]{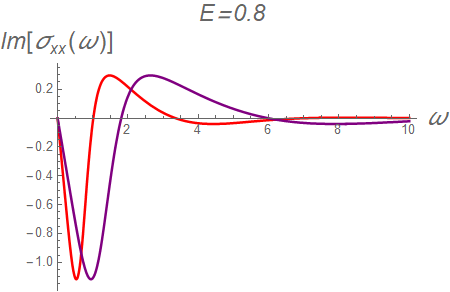}
 	\caption{Imaginary part of the longitudinal $\sigma_{xx}$ conductivity for the case $E < B$ (symmetric phase). \label{fig:Im_cond_EltB}}
 \end{figure}

 As in the previous case, it is possible to obtain the imaginary part of the conductivities using the Kramers-Kronig relations
 \eqref{eq:kramers-Kronig}. However we report the imaginary part of $\sigma_{xx}$ as an example in Figure~\ref{fig:Im_cond_EltB}.

 We find for the imaginary parts a standard behavior, very similar to the one we have obtained for the $E>B$ case. Indeed, they vanish in both the low frequency and high frequency limits and they have finite peaks at intermediate frequencies. We have very similar plots for $\imm[\sigma_{yy}]$, while the imaginary parts of the transverse conductivities vanish for every frequency. This is consistent with the fact that their real parts are just constant.

From the plots in Figure~\ref{fig:real_cond_EltB} we notice that when the electric field is small (\eg $E=0.1$) the Ohm conductivities $\sigma_{xx}$
and $\sigma_{yy}$ are almost equal, while they become clearly different for higher values of the electric field (\eg $E=0.8$). 
This is consistent with the fact that the background electric field $E$ is what actually breaks the rotational symmetry on the 2-dimensional semimetal sheet.

\subsubsection*{Non symmetric phase ($\rho < \sqrt{7}\, B$)}

As we see from the phase diagram in Figure~\ref{fig:phasediag}, when $E<B$ and $\rho<\sqrt{7}\, B$ the system is 
in the chirally broken phase. Therefore in this case we have to deal with background worldvolume configurations for the probe D5
with non-trivial profile along $\psi$. These can be determined by solving (numerically) the equations of motion of the DBI action \eqref{eq:DBI}.

The fact that $\psi$ is not constant makes the computations much more involved. Indeed, when $\psi \neq \pi/2$ the gauge sector does not decouple anymore from the scalar one, as we can see from the action for the fluctuations \eqref{eq:eff_action2}
along with \eqref{eq:coeff_eff_act}. We have then to consider the whole action with scalar fluctuations $\psi^{(1)} (r)$ only along $\psi$, which, to simplify the notation, we denote simply as $\Psi$. Then the effective action for the fluctuations $S_{\rm eff}$ assumes the following expression
\begin{equation}
\label{eq:eff_action_non_chiral}
\begin{split}
S_{\rm eff}=-\CN_5 &\int d^6\sigma  \Biggl\{ \frac{\sqrt{s}}{g_{5}^2}\Bigl[ \mathcal{P}^{ab}_{\psi \psi}\, \Psi_{,a} \Psi_{,b} + \mathcal{Q}_{\psi \psi}\, \Psi^2 +\mathcal{R}_{\psi \psi}^{a} \, \Psi_{,a} \Psi + \frac{1}{4}s^{ac}s^{bd}F^{(1)}_{cb}F^{(1)}_{da}  \\
+& \mathcal{S}_{\psi}^{abc} \Psi_{,a} F^{(1)}_{bc} + \mathcal{T}_{\psi}^{ab}\, \Psi \, F^{(1)}_{ab}\Bigr]+
\frac{1}{8}Q\epsilon^{abcd\theta_1\varphi_1}F^{(1)}_{ab}F^{(1)}_{cd} \Biggr\} \, .
\end{split}
\end{equation}

When the charge density is less than $\sqrt{7}\, B$ but finite the D5-branes have black hole embeddings, namely they do reach 
the Poincar\'e horizon. 
From the $r \to \infty$ limit of the equations of motion derived from this action, we find the following behavior 
for the gauge and scalar fluctuations near the Poincar\'e horizon 
\begin{equation}
\label{behaviour_non_chiral}
\begin{split}
a_i &  = r \, e^{i \,\omega \, B^{3/2} \, r_b^2 \, r} \chi_i (r)  \, ,\\
\Psi & = \, e^{i \,\omega \, B^{3/2} \, r_b^2 \, r} \chi_\Psi (r) \, ,
\end{split}
\end{equation}
where the functions $\chi_i (r)$ and $\chi_\Psi (r)$ admit analytical expansions near the Poincar\'e horizon. 
We shall use these expansions to fix the boundary conditions in  the numerical integration of the equations of motion.

When the charge density vanishes the D5-branes configurations are Minkowski embeddings. In this case the boundary conditions for the
fluctuation fields have to be fixed at the point where the D5-brane worldvolume pinches off. 

\subsubsection*{AC conductivities}

 In the following we show some plots of the conductivities for $E<B$ and $\rho<\sqrt{7}\, B$.
Without loss of generality, the magnetic field $B$ has been set to $1$ in all plots.
We start with the case of finite charge density, \ie $0<\rho<\sqrt{7}\, B$.

\begin{figure}[!h]
	\centering
	\includegraphics[width=.48\textwidth]{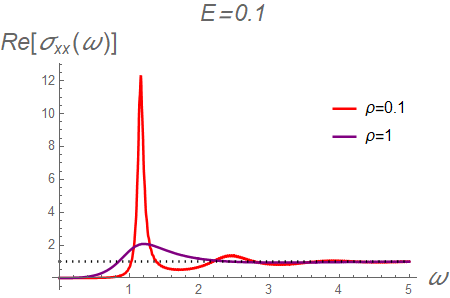}
	\hfill
	\includegraphics[width=.48\textwidth]{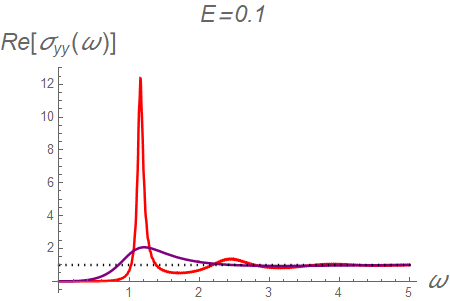}
	\\
	\includegraphics[width=.48\textwidth]{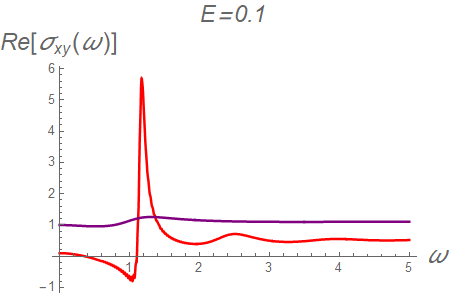}
	\hfill
	\includegraphics[width=.48\textwidth]{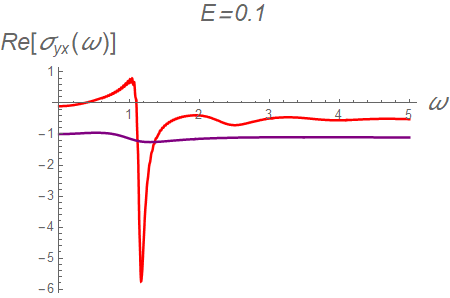}
	\\
	\includegraphics[width=.48\textwidth]{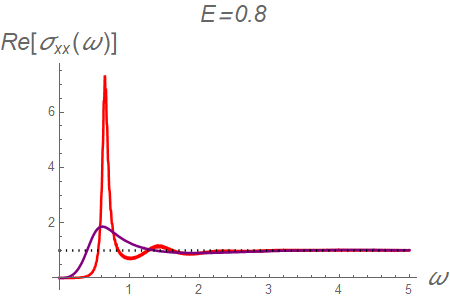}
	\hfill
	\includegraphics[width=.48\textwidth]{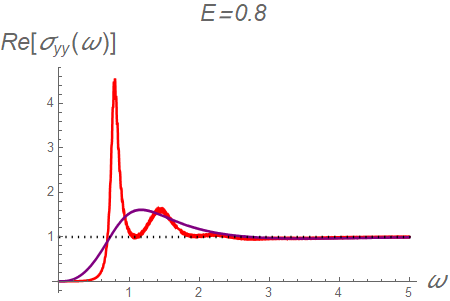}
	\\
	\includegraphics[width=.48\textwidth]{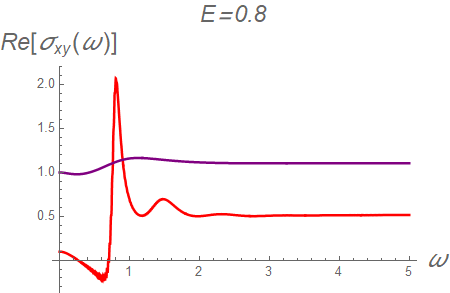}
	\hfill
	\includegraphics[width=.48\textwidth]{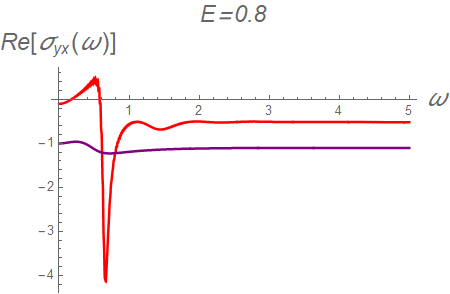}
	\caption{Real parts of the conductivities for different values of electric field $E$ and charge density $\rho$ in the case $E < B$ (non-symmetric phase).\label{fig:real_cond_EltB2}}
\end{figure}

The behavior of the real part of the longitudinal conductivities for small and large frequencies is the same as for the symmetric phase ($E<B$) case. Indeed, they again approach to a constant in the high frequency limit and they go to zero as $\omega \to 0$ consistently with the vanishing DC conductivities. 
At intermediate frequencies we notice the presence of some peaks, which become narrower and higher as the charge density goes to zero. 
For the real parts of the transverse conductivities we observe instead a different behavior with respect to the one seen for the symmetric phase. Indeed in this case they are not just trivially constant, but they vary with the frequency. They start from the DC value, have extremal points for intermediate frequencies and become constant in the high frequency limit. Again the smaller the charge density the higher are the amplitudes of the peaks. 

\begin{figure}[!h]
	\centering
	\includegraphics[width=.48\textwidth]{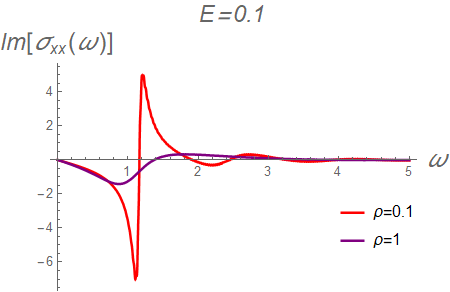}
	\hfill
	\includegraphics[width=.48\textwidth]{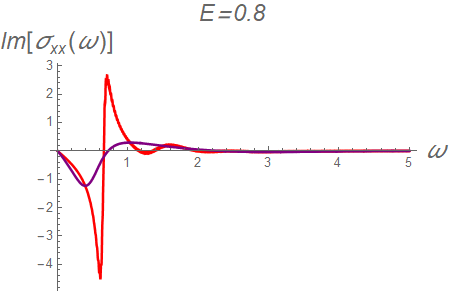}
	\caption{Imaginary part of the longitudinal $\sigma_{xx}$ conductivity in the case $E < B$ (non-symmetric phase). \label{fig:Im_cond_EltB2}}
\end{figure}
For completeness Figure~\ref{fig:Im_cond_EltB2} shows some examples of the imaginary part of the longitudinal $\sigma_{xx}$ conductivity. 
From these plots we observe the same behavior of the symmetric $E<B$ case for these imaginary parts. The same happens for the other longitudinal conductivity $\sigma_{yy}$. For the transverse conductivities we have a different situation with respect to the symmetric case, as it happens for the real parts. Indeed, the imaginary parts are not zero, but they vary with the frequency in a way very similar to the longitudinal case.

Also in this case we see that as the value of the background electric field approaches zero the system tends to recover the 
2-dimensional rotational symmetry: indeed for small $E$ we have $\sigma_{xx}\simeq \sigma_{yy}$ and $\sigma_{xy}\simeq -\sigma_{yx}$.

\begin{figure}[!h]
	\centering
	\includegraphics[scale=.6]{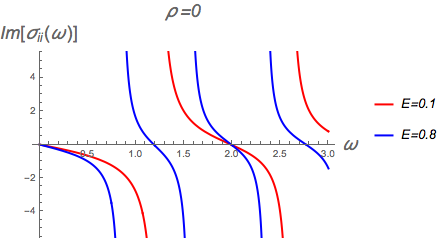}
	\caption{Imaginary part of the longitudinal $\sigma_{xx}=\sigma_{yy}$ conductivities in the case $\rho=0$ (Minkowski embedding), $E < B$. \label{fig:Im_cond_neutral}}
\end{figure}

At zero charge density, where the background configurations for the D5-branes are Minkowski embeddings, all the real parts of the conductivities 
identically vanish, except for the presence of delta function peaks in the longitudinal conductivities, 
that can be identified looking at their imaginary parts. In Figure~\ref{fig:Im_cond_neutral} we show, as examples, two plots of the
imaginary part of the longitudinal conductivity. Note that when $\rho=0$, $\sigma_{xx}=\sigma_{yy}$ and that the Hall conductivities are identically zero, 
so the conductivity matrix still has a rotational symmetry, even in presence of the electric field.

The  behavior of the conductivities in Figure~\ref{fig:Im_cond_neutral} confirms our previous observation that the peaks in 
the real part of the conductivities tend to become delta functions in the zero charge density limit.


\section{Discussion}
\label{sec:discussion}

We used the D3/probe-D5-brane system as a top-down holographic model for a Dirac semimetal like graphene. 
In particular, we considered the system at finite charge density $\rho$ and in the presence of
mutually orthogonal electric and magnetic fields at zero temperature. 
The phase diagram, depicted in Figure~\ref{fig:phasediag}, shows two stable phases for the system:
the one with broken chiral symmetry, favored when $E<B$ and $\rho < \sqrt{7}\, B$,
and the chirally symmetric one, favored in the remainder part of the phase space.
Studying the fluctuations around stable background configurations we were able to compute the AC conductivity matrices
for the system. 

All the conductivities derived in our model have the expected behavior in the small and high frequency regimes.
Indeed in the $\omega\to 0$ limit we recover exactly the DC values that can be obtained using the Karch-O'Bannon method to fix the 
currents. 
In the high frequency limit the real part of the conductivities goes to a constant; 
this is a standard behavior of any (2+1)-dimensional systems where the conductivity is dimensionless. 
The imaginary parts go to zero both in the low and high frequency limits. 

When $E>B$ the system is in the metallic phase. 
The real part of the conductivities stays finite in the low 
frequency regime: this is evident from the plots in Figure~\ref{fig:real_cond_EgtB}, and 
it is also confirmed by the vanishing of the imaginary parts of the conductivities at zero frequency, since a delta function peak would
appear as a divergence in the imaginary part. 
Moreover, it is worth noticing that as $E \to B$, a Drude-like peak does emerge at small frequencies. 
This is the expected behavior in probe brane systems \cite{Hartnoll:2009ns,Das:2010yw,Hartnoll:2016apf} and 
it is due to the fact that in this limit the effective temperature felt by open string excitations tends to zero
and the system approximately recovers the conservation of the charge current operator \cite{Chen:2017dsy}.

It is possible to compare the behavior of the conductivities we found with some experimental measures performed on graphene or similar materials. For example, it is found that for high quality graphene on silicon dioxide substrates, the AC conductivity in the THz frequency range is well described by a classical Drude model \cite{Boggild-2017}. The assumptions behind this model are that there must be an electric field $E$ which accelerates the charge carriers and that the scattering events are instantaneous and isotropic. Under these hypotheses, the conductivity of high quality graphene should look as in Figure 13 (c-e) of \cite{Boggild-2017}. Very similar experimental results have been found also in \cite{Horng-2011} (see Figure 2 reported there).
This experimental picture is compatible with what we found for the AC conductivity in the case $E>B$. In most of our plots, the similarity with the Drude model and with the experimental measurements is striking (even if we start to see deviations when the charge is high).

In the $E < B$,  $\rho>\sqrt{7}\, B$ case we obtained a trivial frequency dependence for the  transverse AC conductivities, which
therefore are fixed only by Lorentz invariance \cite{OBannon:2007cex,Hartnoll:2007ai}.

In the chirally broken phase, namely when $E<B$, $\rho<\sqrt{7}\, B$, a peculiar and particularly interesting behavior in the conductivity does emerge.
From the plots in Figure~\ref{fig:real_cond_EltB2} we clearly notice the presence of some peaks in the conductivity which become sharper as the
charge density decreases and eventually turn into delta functions when $\rho$ is exactly zero.
These peaks can be interpreted as resonances that appear when the system is (almost-)neutral and that are otherwise concealed by the 
presence of the charge density.
These resonances are related to the chiral condensates, which indeed are present only in this region of the phase space. 
It would be worth to further investigate if this interpretation is indeed correct.
If this should be the case, the presence of the peaks would be a remarkable outcome of our model since it 
would show that the effects of the chiral condensates can be observed in the optical
conductivities.


\section*{Acknowledgments}

We would like to thank Francisco Pe\~na Benitez for helpful comments and discussions.

\bibliographystyle{JHEP}

\bibliography{biblio}

\providecommand{\href}[2]{#2}\begingroup\raggedright\begin{thebibliography}{10}

\bibitem{Maldacena:1997re}
J.~M. Maldacena, \emph{{The Large N limit of superconformal field theories and
  supergravity}}, \href{https://doi.org/10.1023/A:1026654312961,
  10.4310/ATMP.1998.v2.n2.a1}{\emph{Int. J. Theor. Phys.} {\bfseries 38} (1999)
  1113} [\href{https://arxiv.org/abs/hep-th/9711200}{{\ttfamily
  hep-th/9711200}}].

\bibitem{Gubser:1998bc}
S.~S. Gubser, I.~R. Klebanov and A.~M. Polyakov, \emph{{Gauge theory
  correlators from noncritical string theory}},
  \href{https://doi.org/10.1016/S0370-2693(98)00377-3}{\emph{Phys. Lett.}
  {\bfseries B428} (1998) 105}
  [\href{https://arxiv.org/abs/hep-th/9802109}{{\ttfamily hep-th/9802109}}].

\bibitem{Witten:1998qj}
E.~Witten, \emph{{Anti-de Sitter space and holography}},
  \href{https://doi.org/10.4310/ATMP.1998.v2.n2.a2}{\emph{Adv. Theor. Math.
  Phys.} {\bfseries 2} (1998) 253}
  [\href{https://arxiv.org/abs/hep-th/9802150}{{\ttfamily hep-th/9802150}}].

\bibitem{Crossno:2016}
J.~{Crossno}, J.~K. {Shi}, K.~{Wang}, X.~{Liu}, A.~{Harzheim}, A.~{Lucas}
  et~al., \emph{{Observation of the Dirac fluid and the breakdown of the
  Wiedemann-Franz law in graphene}},
  \href{https://doi.org/10.1126/science.aad0343}{\emph{Science} {\bfseries 351}
  (2016) 1058} [\href{https://arxiv.org/abs/1509.04713}{{\ttfamily
  1509.04713}}].

\bibitem{b92bfba555b644c5a82fba0fc93cd45b}
R.~V. Gorbachev, A.~K. Geim, M.~I. Katsnelson, K.~S. Novoselov, T.~Tudorovskiy,
  I.~V. Grigorieva et~al., \emph{{Strong Coulomb drag and broken symmetry in
  double-layer graphene}},
  \href{https://doi.org/10.1038/nphys2441}{\emph{Nature Physics} {\bfseries 8}
  (2012) 896}.

\bibitem{Seo:2016vks}
Y.~Seo, G.~Song, P.~Kim, S.~Sachdev and S.-J. Sin, \emph{{Holography of the
  Dirac Fluid in Graphene with two currents}},
  \href{https://doi.org/10.1103/PhysRevLett.118.036601}{\emph{Phys. Rev. Lett.}
  {\bfseries 118} (2017) 036601}
  [\href{https://arxiv.org/abs/1609.03582}{{\ttfamily 1609.03582}}].

\bibitem{Rogatko:2017tae}
M.~Rogatko and K.~I. Wysokinski, \emph{{Two interacting current model of
  holographic Dirac fluid in graphene}},
  \href{https://doi.org/10.1103/PhysRevD.97.024053}{\emph{Phys. Rev.}
  {\bfseries D97} (2018) 024053}
  [\href{https://arxiv.org/abs/1708.08051}{{\ttfamily 1708.08051}}].

\bibitem{Rogatko:2017svr}
M.~Rogatko and K.~I. Wysokinski, \emph{{Holographic calculation of the
  magneto-transport coefficients in Dirac semimetals}},
  \href{https://doi.org/10.1007/JHEP01(2018)078}{\emph{JHEP} {\bfseries 01}
  (2018) 078} [\href{https://arxiv.org/abs/1712.01608}{{\ttfamily
  1712.01608}}].

\bibitem{DeWolfe:2001pq}
O.~DeWolfe, D.~Z. Freedman and H.~Ooguri, \emph{{Holography and defect
  conformal field theories}},
  \href{https://doi.org/10.1103/PhysRevD.66.025009}{\emph{Phys. Rev.}
  {\bfseries D66} (2002) 025009}
  [\href{https://arxiv.org/abs/hep-th/0111135}{{\ttfamily hep-th/0111135}}].

\bibitem{Karch:2000gx}
A.~Karch and L.~Randall, \emph{{Open and closed string interpretation of SUSY
  CFT's on branes with boundaries}},
  \href{https://doi.org/10.1088/1126-6708/2001/06/063}{\emph{JHEP} {\bfseries
  06} (2001) 063} [\href{https://arxiv.org/abs/hep-th/0105132}{{\ttfamily
  hep-th/0105132}}].

\bibitem{Erdmenger:2002ex}
J.~Erdmenger, Z.~Guralnik and I.~Kirsch, \emph{{Four-dimensional superconformal
  theories with interacting boundaries or defects}},
  \href{https://doi.org/10.1103/PhysRevD.66.025020}{\emph{Phys. Rev.}
  {\bfseries D66} (2002) 025020}
  [\href{https://arxiv.org/abs/hep-th/0203020}{{\ttfamily hep-th/0203020}}].

\bibitem{Filev:2007gb}
V.~G. Filev, C.~V. Johnson, R.~C. Rashkov and K.~S. Viswanathan,
  \emph{{Flavoured large N gauge theory in an external magnetic field}},
  \href{https://doi.org/10.1088/1126-6708/2007/10/019}{\emph{JHEP} {\bfseries
  10} (2007) 019} [\href{https://arxiv.org/abs/hep-th/0701001}{{\ttfamily
  hep-th/0701001}}].

\bibitem{Filev:2009xp}
V.~G. Filev, C.~V. Johnson and J.~P. Shock, \emph{{Universal Holographic Chiral
  Dynamics in an External Magnetic Field}},
  \href{https://doi.org/10.1088/1126-6708/2009/08/013}{\emph{JHEP} {\bfseries
  08} (2009) 013} [\href{https://arxiv.org/abs/0903.5345}{{\ttfamily
  0903.5345}}].

\bibitem{Evans:2010hi}
N.~Evans, A.~Gebauer, K.-Y. Kim and M.~Magou, \emph{{Phase diagram of the D3/D5
  system in a magnetic field and a BKT transition}},
  \href{https://doi.org/10.1016/j.physletb.2011.03.004}{\emph{Phys. Lett.}
  {\bfseries B698} (2011) 91}
  [\href{https://arxiv.org/abs/1003.2694}{{\ttfamily 1003.2694}}].

\bibitem{Kaplan:2009kr}
D.~B. Kaplan, J.-W. Lee, D.~T. Son and M.~A. Stephanov, \emph{{Conformality
  Lost}}, \href{https://doi.org/10.1103/PhysRevD.80.125005}{\emph{Phys. Rev.}
  {\bfseries D80} (2009) 125005}
  [\href{https://arxiv.org/abs/0905.4752}{{\ttfamily 0905.4752}}].

\bibitem{Jensen:2010ga}
K.~Jensen, A.~Karch, D.~T. Son and E.~G. Thompson, \emph{{Holographic
  Berezinskii-Kosterlitz-Thouless Transitions}},
  \href{https://doi.org/10.1103/PhysRevLett.105.041601}{\emph{Phys. Rev. Lett.}
  {\bfseries 105} (2010) 041601}
  [\href{https://arxiv.org/abs/1002.3159}{{\ttfamily 1002.3159}}].

\bibitem{Evans:2013jma}
N.~Evans and K.-Y. Kim, \emph{{Vacuum alignment and phase structure of
  holographic bi-layers}},
  \href{https://doi.org/10.1016/j.physletb.2013.11.060}{\emph{Phys. Lett.}
  {\bfseries B728} (2014) 658}
  [\href{https://arxiv.org/abs/1311.0149}{{\ttfamily 1311.0149}}].

\bibitem{Grignani:2014vaa}
G.~Grignani, N.~Kim, A.~Marini and G.~W. Semenoff, \emph{{Holographic
  D3-probe-D5 Model of a Double Layer Dirac Semimetal}},
  \href{https://doi.org/10.1007/JHEP12(2014)091}{\emph{JHEP} {\bfseries 12}
  (2014) 091} [\href{https://arxiv.org/abs/1410.4911}{{\ttfamily 1410.4911}}].

\bibitem{Grignani:2016npu}
G.~Grignani, A.~Marini, A.~C. Pigna and G.~W. Semenoff, \emph{{Phase structure
  of a holographic double monolayer Dirac semimetal}},
  \href{https://doi.org/10.1007/JHEP06(2016)141}{\emph{JHEP} {\bfseries 06}
  (2016) 141} [\href{https://arxiv.org/abs/1603.02583}{{\ttfamily
  1603.02583}}].

\bibitem{Karch:2007pd}
A.~Karch and A.~O'Bannon, \emph{{Metallic AdS/CFT}},
  \href{https://doi.org/10.1088/1126-6708/2007/09/024}{\emph{JHEP} {\bfseries
  09} (2007) 024} [\href{https://arxiv.org/abs/0705.3870}{{\ttfamily
  0705.3870}}].

\bibitem{Evans:2011tk}
N.~Evans, K.-Y. Kim, J.~P. Shock and J.~P. Shock, \emph{{Chiral phase
  transitions and quantum critical points of the D3/D7(D5) system with mutually
  perpendicular E and B fields at finite temperature and density}},
  \href{https://doi.org/10.1007/JHEP09(2011)021}{\emph{JHEP} {\bfseries 09}
  (2011) 021} [\href{https://arxiv.org/abs/1107.5053}{{\ttfamily 1107.5053}}].

\bibitem{Hartnoll:2009ns}
S.~A. Hartnoll, J.~Polchinski, E.~Silverstein and D.~Tong, \emph{{Towards
  strange metallic holography}},
  \href{https://doi.org/10.1007/JHEP04(2010)120}{\emph{JHEP} {\bfseries 04}
  (2010) 120} [\href{https://arxiv.org/abs/0912.1061}{{\ttfamily 0912.1061}}].

\bibitem{Das:2010yw}
S.~R. Das, T.~Nishioka and T.~Takayanagi, \emph{{Probe Branes, Time-dependent
  Couplings and Thermalization in AdS/CFT}},
  \href{https://doi.org/10.1007/JHEP07(2010)071}{\emph{JHEP} {\bfseries 07}
  (2010) 071} [\href{https://arxiv.org/abs/1005.3348}{{\ttfamily 1005.3348}}].

\bibitem{OBannon:2007cex}
A.~O'Bannon, \emph{{Hall Conductivity of Flavor Fields from AdS/CFT}},
  \href{https://doi.org/10.1103/PhysRevD.76.086007}{\emph{Phys. Rev.}
  {\bfseries D76} (2007) 086007}
  [\href{https://arxiv.org/abs/0708.1994}{{\ttfamily 0708.1994}}].

\bibitem{Mateos:2006nu}
D.~Mateos, R.~C. Myers and R.~M. Thomson, \emph{{Holographic phase transitions
  with fundamental matter}},
  \href{https://doi.org/10.1103/PhysRevLett.97.091601}{\emph{Phys. Rev. Lett.}
  {\bfseries 97} (2006) 091601}
  [\href{https://arxiv.org/abs/hep-th/0605046}{{\ttfamily hep-th/0605046}}].

\bibitem{Kobayashi:2006sb}
S.~Kobayashi, D.~Mateos, S.~Matsuura, R.~C. Myers and R.~M. Thomson,
  \emph{{Holographic phase transitions at finite baryon density}},
  \href{https://doi.org/10.1088/1126-6708/2007/02/016}{\emph{JHEP} {\bfseries
  02} (2007) 016} [\href{https://arxiv.org/abs/hep-th/0611099}{{\ttfamily
  hep-th/0611099}}].

\bibitem{Mas:2008jz}
J.~Mas, J.~P. Shock, J.~Tarrio and D.~Zoakos, \emph{{Holographic Spectral
  Functions at Finite Baryon Density}},
  \href{https://doi.org/10.1088/1126-6708/2008/09/009}{\emph{JHEP} {\bfseries
  09} (2008) 009} [\href{https://arxiv.org/abs/0805.2601}{{\ttfamily
  0805.2601}}].

\bibitem{Kim:2011qh}
K.-Y. Kim, J.~P. Shock and J.~Tarrio, \emph{{The open string membrane paradigm
  with external electromagnetic fields}},
  \href{https://doi.org/10.1007/JHEP06(2011)017}{\emph{JHEP} {\bfseries 06}
  (2011) 017} [\href{https://arxiv.org/abs/1103.4581}{{\ttfamily 1103.4581}}].

\bibitem{Seiberg:1999vs}
N.~Seiberg and E.~Witten, \emph{{String theory and noncommutative geometry}},
  \href{https://doi.org/10.1088/1126-6708/1999/09/032}{\emph{JHEP} {\bfseries
  09} (1999) 032} [\href{https://arxiv.org/abs/hep-th/9908142}{{\ttfamily
  hep-th/9908142}}].

\bibitem{Gibbons:2000xe}
G.~W. Gibbons and C.~A.~R. Herdeiro, \emph{{Born-Infeld theory and stringy
  causality}}, \href{https://doi.org/10.1103/PhysRevD.63.064006}{\emph{Phys.
  Rev.} {\bfseries D63} (2001) 064006}
  [\href{https://arxiv.org/abs/hep-th/0008052}{{\ttfamily hep-th/0008052}}].

\bibitem{Ryu:2011vq}
S.~Ryu, T.~Takayanagi and T.~Ugajin, \emph{{Holographic Conductivity in
  Disordered Systems}},
  \href{https://doi.org/10.1007/JHEP04(2011)115}{\emph{JHEP} {\bfseries 04}
  (2011) 115} [\href{https://arxiv.org/abs/1103.6068}{{\ttfamily 1103.6068}}].

\bibitem{Chen:2017dsy}
C.-F. Chen and A.~Lucas, \emph{{Origin of the Drude peak and of zero sound in
  probe brane holography}},
  \href{https://doi.org/10.1016/j.physletb.2017.10.023}{\emph{Phys. Lett.}
  {\bfseries B774} (2017) 569}
  [\href{https://arxiv.org/abs/1709.01520}{{\ttfamily 1709.01520}}].

\bibitem{Grozdanov:2018fic}
S.~Grozdanov, A.~Lucas and N.~Poovuttikul, \emph{{Holography and hydrodynamics
  with weakly broken symmetries}},
  \href{https://arxiv.org/abs/1810.10016}{{\ttfamily 1810.10016}}.

\bibitem{Hartnoll:2016apf}
S.~A. Hartnoll, A.~Lucas and S.~Sachdev, \emph{{Holographic quantum matter}},
  \href{https://arxiv.org/abs/1612.07324}{{\ttfamily 1612.07324}}.

\bibitem{Boggild-2017}
P.~B{\o}ggild, D.~M.~A. Mackenzie, P.~R. Whelan, D.~H. Petersen, J.~D. Buron,
  A.~Zurutuza et~al., \emph{{Mapping the electrical properties of large-area
  graphene}}, \href{https://doi.org/10.1088/2053-1583/aa8683}{\emph{2D
  Materials} {\bfseries 4} (2017) 042003}.

\bibitem{Horng-2011}
J.~Horng, C.-F. Chen, B.~Geng, C.~Girit, Y.~Zhang, Z.~Hao et~al., \emph{{Drude
  conductivity of Dirac fermions in graphene}},
  \href{https://doi.org/10.1103/PhysRevB.83.165113}{\emph{Physical Review B}
  {\bfseries 83} (2011) 165113}.

\bibitem{Hartnoll:2007ai}
S.~A. Hartnoll and P.~Kovtun, \emph{{Hall conductivity from dyonic black
  holes}}, \href{https://doi.org/10.1103/PhysRevD.76.066001}{\emph{Phys. Rev.}
  {\bfseries D76} (2007) 066001}
  [\href{https://arxiv.org/abs/0704.1160}{{\ttfamily 0704.1160}}].

\end{thebibliography}\endgroup


\end{document}